\documentclass{jfm}
\usepackage{graphicx}
\usepackage{epstopdf, epsfig}
\usepackage{amsfonts,amssymb,amsmath,mathbbol}   % For maths symbols.

\usepackage{color} %new package
\shorttitle{Effect of swirl air on droplet morphology}
\shortauthor{P. K. Kirar, S. K. Soni, P. S. Kolhe and K. C. Sahu}

\title{An experimental investigation of droplet morphology in swirl flow}

\author{Pavan Kumar Kirar,\aff{1}
  Surendra Kumar Soni,\aff{2}
  Pankaj S. Kolhe\aff{2}$\dagger$
 \and Kirti Chandra Sahu\aff{1}\corresp{\email{psk@mae.iith.ac.in, ksahu@che.iith.ac.in}}}

\affiliation{\aff{1}Department of Chemical Engineering, Indian Institute of Technology Hyderabad, Sangareddy, 502 284, Telangana, India
\aff{2}Department of Mechanical and Aerospace Engineering, Indian Institute of Technology Hyderabad, Sangareddy, 502 284, Telangana, India}

\begin{document}

\maketitle

\begin{abstract}
The interaction of a droplet with a swirling airstream is investigated experimentally by shadowgraphy and particle image velocimetry techniques. In swirl flow, the droplet experiences oppose-flow, cross-flow, and co-flow conditions depending on its ejection location, the velocity of the airstream, and swirl strength, which results in distinct droplet morphologies as compared to the straight airflow situation. We observe a new breakup phenomenon, termed as `retracting bag breakup', as the droplet encounters a differential flow field created by the wake of the swirler's vanes and the central recirculation zone in swirl airflow. A regime map demarcating the various modes, such as no breakup, vibrational breakup, retracting bag breakup, and bag breakup modes, is presented for different sets of dimensionless parameters influencing the droplet morphology and its trajectory. In contrast to the straight flow, the swirl flow promotes the development of the Rayleigh-Taylor instability, enhancing the stretching factor in the droplet deformation process, resulting in a larger number of fingers on the droplet's surface. In order to gain physical insight, a modified theoretical analysis based on the Rayleigh-Taylor instability is proposed for the swirl flow. The experimental behaviour of droplet deformation phenomena in swirl flow conditions can be determined by modifying the stretching factor in the theoretical model.
\end{abstract}


%%%%%%%%%%%%%%%%%%%%%%%%%%%%%%%%%%%%%%%%%%%%%%%%%%%%%%%%%
%%%%%%%%%%%%%%%%%%%%%%%%%%%%%%%%%%%%%%%%%%%%%%%%%%%%%%%%%

\section{Introduction} \label{sec:intro}
When liquid droplets are exposed to an airstream, they experience morphological changes as a result of Rayleigh–Taylor inertial instability during the initial flattening stage and subsequent break up as a consequence of Rayleigh–Plateau capillary instability \citep{taylor1963shape,jackiw2021aerodynamic}. This phenomenon is observed in many industrial applications, e.g. fuel atomization and combustion \citep{reitz1987structure,lefebvre2017atomization} and natural phenomena \citep{Villermaux2009single,prabhakaran2017can,balla2019shape}. Atomization also aids in the understanding of the migration and spreading of respiratory droplets, as in the case of COVID-19 \citep{mittal2020flow,Katre2021PoF}. Researchers have been studying the fragmentation of a liquid droplet into tiny satellite droplets in a high-speed continuous airstream for decades, taking into consideration in-line \citep{taylor1963shape,komabayasi1964life,dai2001temporal,Villermaux2009single,nykteri2021droplet,flock2012experimental}, cross-flow \citep{pilch1987use,zhao2010morphological,kekesi2014drop,kulkarni2014bag,xiao2017simulation,yang2017transitions,jain2019secondary,fakhari2011investigation}, and oblique \citep{soni2020deformation} configurations. In a cross-flow configuration, a freely falling droplet interacts with the airstream in a direction orthogonal to gravity, whereas in a co-flow/oppose-flow configuration, they interact inline with gravity.

Previous experimental investigations observed that increasing the Weber number ($We \equiv \rho_a U^2 d_0/\sigma_l$, i.e. the inertia force over the surface tension force) causes numerous modes, such as the vibrational, bag, bag-stamen, multi-bag, shear, and catastrophic breakup modes \citep{pilch1987use,dai2001temporal,cao2007new,guildenbecher2009secondary,suryaprakash2019secondary,soni2020deformation}. Here, $\rho_a$, $\sigma_l$, $U$ and $d_0$ denote the density of the air, interfacial tension, average velocity of the airstream and equivalent spherical diameter of the droplet, respectively. At low Weber numbers, a droplet exhibits shape oscillations at a certain frequency, which is known as the vibrational mode. As the Weber number increases, the droplet forms a single bag on the leeward side, which is surrounded by a thick liquid rim. Subsequently, very small and larger droplets are produced as a result of fragmentation of the bag and rim, respectively (bag breakup). The bag-stamen and multi-bag morphologies are similar to the bag breakup mode but characterized by a stamen formation at the centre of the drop, resulting in a large additional drop during the breakup (in bag-stamen mode) and multiple bags formation throughout the drop (in multi-bag mode). The drop's periphery is deflected downstream in shear mode, generating a sheet that breaks into small droplets. High Weber numbers cause the catastrophic breakup mode, in which the droplet explodes into a cluster of fragments very quickly. 

\begin{table}
	\centering
	\begin{tabular}{ccc}
		Reference    &       $We_{cr}$      & Configuration  \\  \hline
	    \cite{pilch1987use}   & $12$ &  Cross-flow \\  
		\cite{guildenbecher2009secondary}  & $11$  &   Cross-flow \\  
		\cite{krzeczkowski1980measurement}  &  $10$ &   Cross-flow \\  
		\cite{Jain2015}   &  $12$  &   Cross-flow \\  
		\cite{hsiang1993drop}   &  $11 \pm 2$  & Cross-flow  \\ 
		\cite{wierzba1990deformation}    &  $13.7$ & Cross-flow  \\  
		\cite{kulkarni2014bag}   & $12$ & Cross-flow  \\  
		\cite{wang2014}  & $10$  &  Cross-flow  \\  
		\cite{Villermaux2009single,villermaux2011distribution}  & $6$  &  Oppose-flow  \\  
		\cite{soni2020deformation}  & $6-12$  &  Oblique  \\ 
	\end{tabular}
         \caption{The critical Weber numbers (transition from vibrational to bag breakup) in various configurations, as reported in earlier investigations.} 
	\label{T:wecr} 
\end{table}

Since the pioneering work of \cite{taylor1963shape}, many researchers have identified the critical Weber number $(We_{cr})$ at which the transition from the vibrational to bag breakup occurs in various configurations. A few key contributions in this subject are listed in Table \ref{T:wecr}. It can be observed that the critical Weber numbers in cross-flow and oppose-flow configurations are around 12 and 6, respectively. \cite{soni2020deformation} investigated the interaction of the continuous airstream at different orientations with a droplet freely falling under gravity and found that the value of $We_{cr}$ decreases as the orientation shifts from the cross-flow to the oppose-flow configuration and asymptotically approaches $We_{cr} \approx 6$ for an angle of inclination of the airstream with the horizontal, $\alpha>60^\circ$. They also observed that the droplet in an oblique configuration exhibits a curvilinear motion while undergoing topological changes. The critical Weber number was also found to be dependent on the initial droplet size, fluid properties of the liquid, ejection height from the nozzle, and velocity profile/potential core region \citep{hanson1963,wierzba1990deformation}. These findings motivate us to investigate the effect of a swirl flow on droplet morphology, which has not yet been studied to our knowledge despite being observed in a variety of situations, such as falling raindrops. A few researchers (e.g. \cite{merkle2003effect,rajamanickam2017dynamics,kumar2019large,patil2021air,soni2021liquid}), however, have examined the characteristics of swirl flow in the absence of droplets using high-speed imaging and particle image velocimetry (PIV) techniques.

The present study investigates the interaction of a swirl flow with an ethanol droplet in a cross-flow configuration. Earlier, the droplet breakup phenomenon in various configurations has been studied using the shock tube method \citep{hsiang1993drop,dai2001temporal,krzeczkowski1980measurement}, the continuous air jet method \citep{kulkarni2014bag,Jain2015,soni2020deformation} and the droplet tower method \citep{Villermaux2009single}. A continuous air jet method with a mechanism to produce swirl airflow has been implemented in our experiment. The droplet's time to cross the shear boundary layer is made significantly shorter than the droplet's resident time when it deforms and breaks in order to ensure that the droplet interacts in the potential core region of the continuous airstream. A shadowgraphy technique using two high-speed imaging systems is employed to record the three-dimensional trajectory and breakup morphology of the droplet for different values of swirl strength and Weber number. The flow field due to the imposed swirl is analyzed using the particle image velocimetry (PIV) technique. A new breakup mode, termed as `retracting bag breakup', is observed in some swirl flow conditions. In contrast to a drop undergoing convectional breakup modes, a deformed disk-shaped drop experiences a differential flow field due to the wake of the vanes and the recirculation zone in a swirl airstream. Subsequently, the drop creates a bag that retracts in the upper half while remaining intact in the lower half, causing the retreating bag breakup phenomenon. A regime map demarcating the various modes is also presented. A theoretical analysis based on the Rayleigh–Taylor instability is established for the swirl flow and compared with our experimental results by analyzing the number of fingers generated around the droplet's rim at the onset of bag breakup. 

The rest of the paper is organized as follows. In \S\ref{sec:expt}, the experimental set-up and procedure are elaborated. The experimental findings are presented in \S\ref{sec:dis}, wherein the various morphologies and trajectories of the droplet are thoroughly analyzed. The characteristics of the swirl airstream as assessed using particle image velocimetry are also shown. Finally, a theoretical model is developed and compared with the experimental results. The concluding remarks are given in \S\ref{sec:conc}.

\section{Experimental set-up}
\label{sec:expt}

\begin{figure}
\centering
\hspace{1.4cm} (a)   \hspace{5.5cm}  (b) \\
\vspace{0.4cm} 
\includegraphics[height=0.5\textwidth]{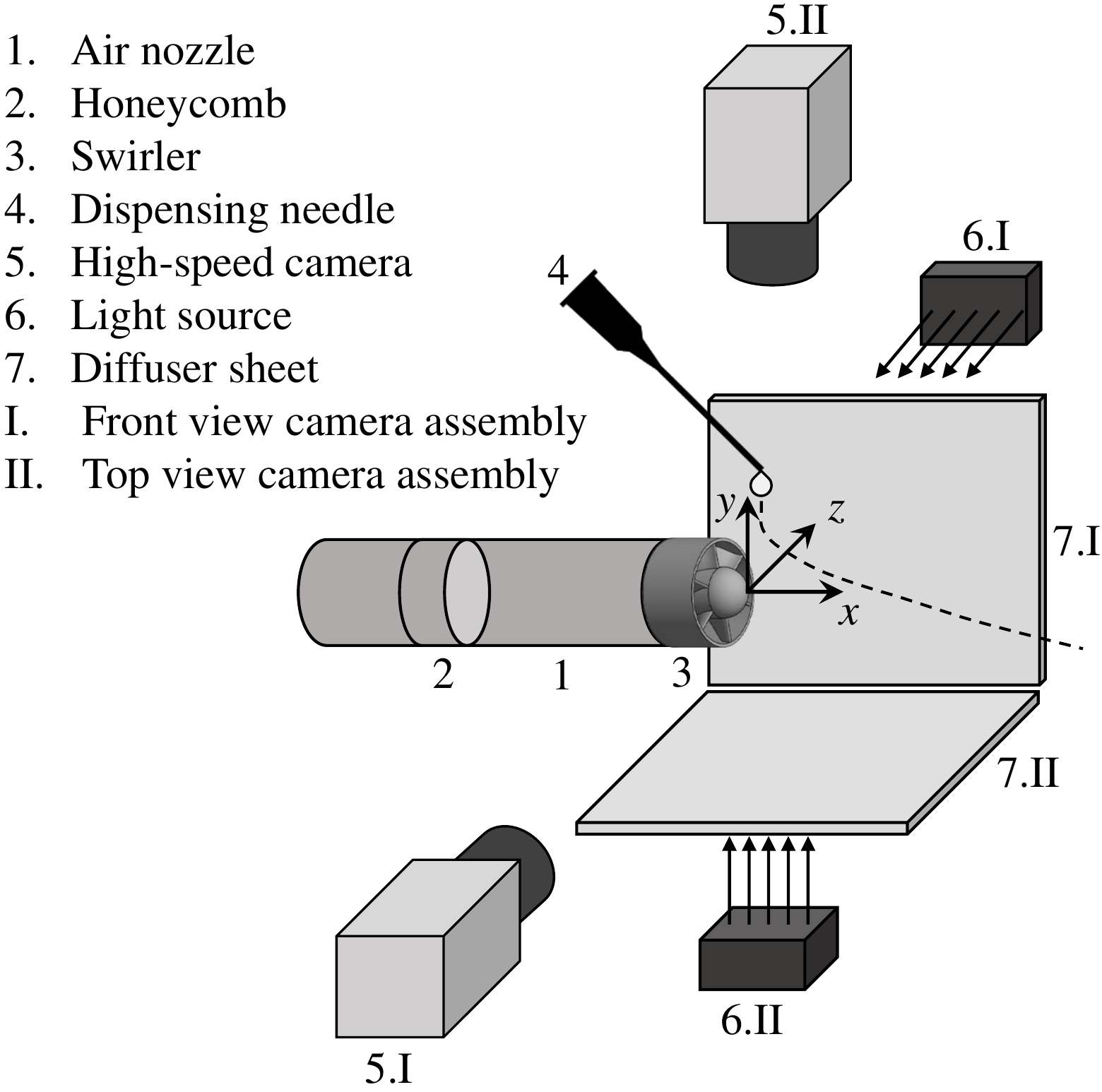} \hspace{1.0cm} \includegraphics[height=0.5\textwidth]{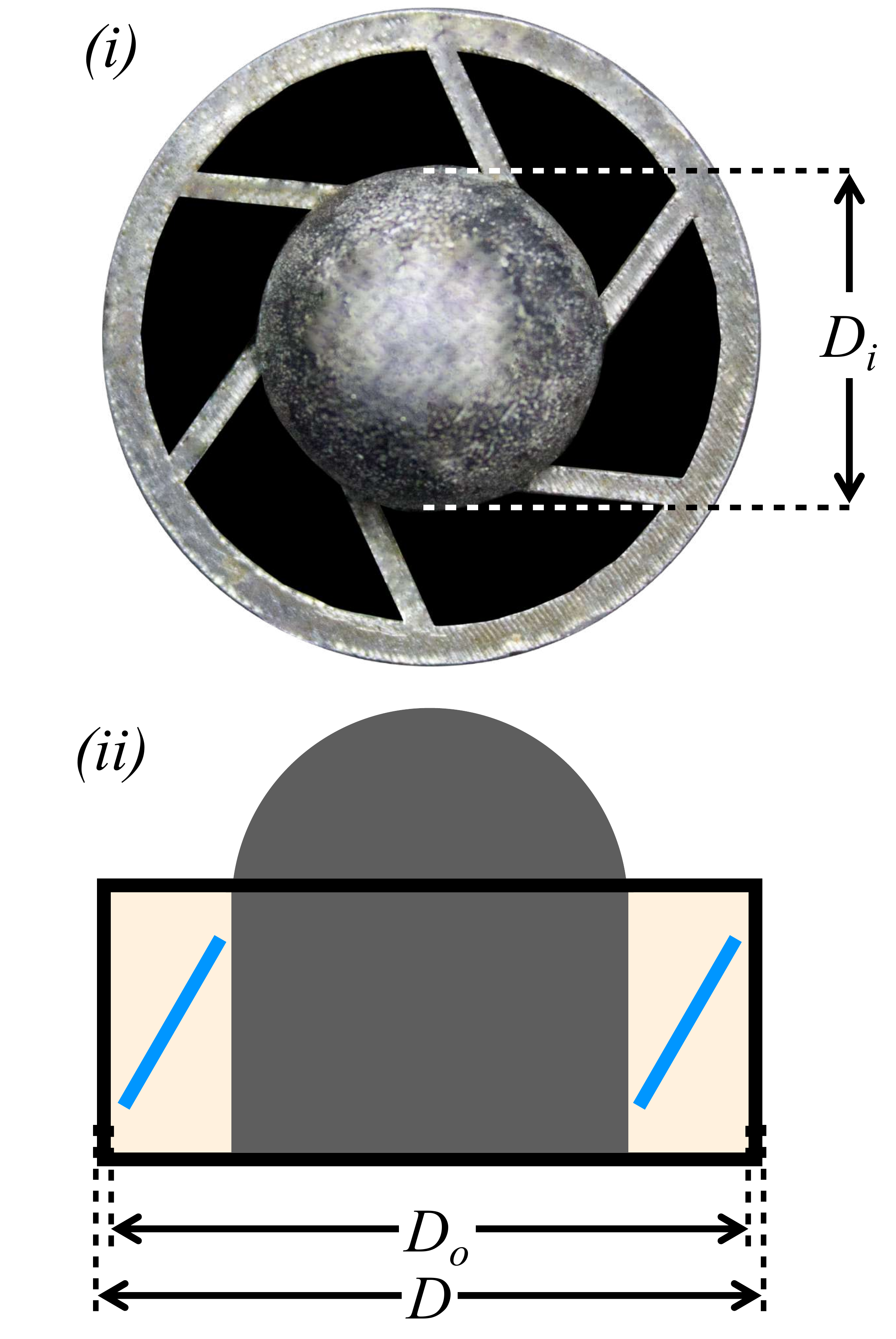}\\
\caption{(a) Schematic diagram of the experimental set-up. It consists of a two-directional image acquisition system involving two high-speed cameras and a set of diffuser sheet and light source for each camera, an air nozzle with a swirler, and a droplet dispensing needle to generate the droplet. (b) The side $(y-z)$ and front $(x-y)$ views of the swirler are shown in panels $(i)$ and $(ii)$, respectively. Two swirlers with different vane angles but the same dimension ($D=23$ mm, $D_o=20$ mm and $D_i=12$ mm) are used in the present study.}
\label{fig1}
\end{figure}

The schematic diagram of the experimental set-up is depicted in figure \ref{fig1}(a). It consists of (i) two high-speed cameras, each with diffused backlit illumination (two-directional image acquisition system), (ii) an air nozzle with a swirler, (iii) a droplet dispensing needle connected with a three-dimensional (3D) traverse system and (iv) a data acquisition system. To achieve low and high swirl strength, two types of swirlers (fabricated using EOS maraging steel MS1 using 3D metallic printing technology) having six wings with different vane angles ($\theta=30^\circ$ and $45^\circ$) but the same outer diameter ($D_o=20$ mm), inner diameter ($D_i=12$ mm) and blade thickness (1 mm) are used. The swirl strength is characterized by the Swirl number $(Sw)$, which is defined as $Sw= \frac{2}{3} \left( \frac{ 1- \left (D_i/D_o \right)^3 } {1- \left (D_i/D_o \right)^2} \right)\tan\theta$, wherein $\theta$ is the vane angle of the swirlers \citep{beer1974combustion}. 

The Cartesian coordinate system $(x,y,z)$ with its origin located at the center of the swirler tip as shown in figure \ref{fig1}(a) is used to describe the results. Figure \ref{fig1}(b) shows the side $(y-z)$ and front $(x-y)$ views of a swirler. A metallic circular nozzle with a swirler near the exit of the nozzle is used to generate a swirl flow field. A honeycomb pallet is fitted at upstream of the nozzle to straighten the flow and reduce inlet airstream disturbances. The air nozzle is connected to an ALICAT digital mass flow controller (model: MCR-500SLPM-D/CM, Make: Alicat Scientific, Inc., USA) that can control flow rates from 0 to 500 standard liters per minute (SLPM). The flow meter has an accuracy of about 0.8\% of the reading + 0.2\% of the full scale. For air supply, the mass flow controller is connected to an air compressor. To ensure dry air during the experiment, an air dryer and moisture remover are installed in the compressed airline. The system is set up in a cross-flow configuration.

An ethanol droplet (diameter, $d_0= 2.7 \pm 0.07$ mm) is injected from the dispensing needle once the swirl flow has been fully established. The location of the tip of the dispensing needle is at $(x_d,y_d,z_d)$, which is varied using a 3D traverse system. In the dimensionless form, the location of the dispensing needle is given by $({\bar x_d},{\bar y_d},{\bar z_d}) = (x_d/D_o,y_d/D_o,z_d/D_o)$. In order to generate the same size of liquid droplets, the ethanol flow rate in the dispensing needle is controlled using a Holmarc syringe pump (model: HO-SPLF-2D, Make: Holmarc Opto-Mechatronics Pvt. Ltd., India). Under the influence of gravity, a droplet generated at the tip of a blunt syringe needle detaches from the needle and enters the airstream. Throughout the study, the droplet dispensing needle diameter and liquid flow rate in the syringe pump are kept constant at 20 gauge and 20 $\mu$l/sec. The flow rate is kept low enough such that the droplets are only detached from the needle due to gravity. While the first droplet of interest interacts with the swirl flow airstream, we make sure no additional droplets emerge from the needle. 

\begin{table}
\begin{center}
\begin{tabular}{cccc}
Working fluids & Density (kg/m$^3$)  & Dynamic viscosity (mPa$\cdot$s) &  Surface tension (mN/m) \\ \hline
Ethanol           & ${\rho_l} = 785.8$           & $\mu_l=1.1~~$           & $\sigma_l=22.1$                \\  
Olive oil    & ${\rho_o} =907.9$        & $\mu_o=74.1$  & $\sigma_o=31.9$       \\  
\end{tabular}
\end{center} 
\caption{Properties of fluids of the working fluids.} \label{T1}
\end{table}

For high-speed visualization, two high-speed cameras (model: Phantom VEO 640L, make: Vision research, USA) with Nikkor lens AF Nikkor 50mm f/1.8D are used to capture the front ($x-y$) and top ($x-z$) views of the droplet interaction with the swirl airstream. The shadowgraph technique used in the present study employs two diffused backlit illuminations (GSVITEC light source, model: MultiLED QT, Make: GSVITEC, Germany) to illuminate the background for both cameras placed orthogonally as shown in figure \ref{fig1}(a). We utilized high-power (150 W and 12,000 lumen) light-emitting diode (LED) lights diffused by a diffuser sheet as the background light source to ensure appropriate brightness when employing the narrow aperture. The high-speed cameras are synchronized with LaserPulse synchronizer (model: 610036, Make: TSI, USA). In all our experiments, the resolution of images captured using the high-speed cameras is $1024 \times 640$ pixels and the images are recorded at 6420 frames per second (fps) with an exposure time of 150 $\mu$s and spatial resolution of 72.73 $\mu$m/pixel. The image sequence of the droplet is recorded in the internal memory of the high-speed cameras and then transferred to a high-end computer (make: Dell) for further analysis. The image processing is performed using the MATLAB\textsuperscript{\textregistered} software. Furthermore, a droplet tracking code has been developed that uses a frame subtraction technique to determine the trajectory of the droplet until it breaks up. The results obtained from this code have been validated against the trajectory of the droplet obtained manually for a few typical cases. 

In addition to the shadowgraph technique for droplet dynamics, we also use a stereo-PIV to analyze the swirl flow field. The schematic diagram of the stereo-PIV is presented in figure \ref{figS1} in Appendix. In the stereo-PIV, olive oil is used to seed the air. The properties of ethanol and olive oil (working fluids) are given in Table \ref{T1}. The values of the density ($\rho_{a}$) and dynamic viscosity ($\mu_{a}$) of the surrounding medium (air phase) are $1.2$ kg/m$^3$ and $0.018$ mPa$\cdot$s. As previously discussed, we can generate two values of the Swirl numbers by using two swirlers with different vane angles, namely $Sw = 0.47$ (low swirl strength) and $Sw = 0.82$ (high swirl strength). In the present study, the Weber number $(We \equiv \rho_a {U}^2 d_0/\sigma_l)$ is defined using the resultant velocity $U$ of the swirl airflow obtained from the stereo-PIV. The droplet dynamics in the swirl flow is also compared with the no swirl case. The dispensing needle location is an important parameter to explore since the aerodynamic force is a strong function of spatial location. Thus, the breakup mode can alter depending on where the dispensing needle is located. Table \ref{T2} presents the dimensionless parameters as well as the location of the dispensing needle normalized with the outer diameter of the swirler ($D_o=20$ mm) considered in the present study. To ensure repeatability, we conduct five measurements of the same set of parameters and perform an uncertainty analysis.

The characteristics of the imposed swirl flow are investigated using a stereo-PIV set-up as shown in figure \ref{figS1}. A single jet oil droplet generator (model: 9307, Make: TSI, USA) is employed for seeding in the airstream coming out from the air nozzle (figure \ref{figS1}). The oil droplet generator works based on the Laskin nozzle principle. It has been used to continually inject seeding particles (olive oil droplets) into the airflow in the upstream. The particle mists are about 1 $\mu$m in diameter with good fluidity and stability in airflow. The particle concentration is changed by adjusting the droplet generator's boost pressure, and then it has been kept constant throughout the study. Then the flow field is obtained by properly seeding the flow region. The high-speed cameras are mounted at a $42^\circ$ angle to each other, and they are on one side of the laser sheet. The Rokinon 135 mm F2.0 lenses are used to increase the pixel density. Both cameras are correctly aligned so that the image plane ($x-y$), lens plane, and sheet plane all cross at the same line. This is accomplished by utilizing scheimpflug between the lens and the camera sensor, resulting in a perfectly focused image plan. To lit up the seeded particles, we use a high repetition rate DPSS Nd:YLF laser (model: LD30-527, make: Litron Lasers, UK) and the time interval between the two laser pulses is maintained at 10 $\mu$s. A LaserPulse light arm (model: 610015, make: TSI, USA) is used to direct the laser path. The divergent sheet optics with -15 mm focal-length cylindrical lens with spherical light sheet optics (model: 610026 collimator, make: TSI, USA) at the exit of the light arm is used to create a thin sheet of thickness $1.5$ mm (approximately) to illuminate the field of view. The laser and cameras are synchronized using the LaserPulse synchronizer and operated by the Insight 4G software (version 11.2 TSI Inc.). We first set the flow rate in the mass flow controller, then turned on the laser and captured the flow field images. 

The frame-straddled images are analyzed using Insight 4G software with the cross-correlated sub-regions of image pairs to generate the velocity vectors. The initial interrogation window size of $256 \times 128$ pixels and final interrogation window size of $64 \times 64$ pixels with 70\% overlap is used to calculate the velocity vectors. The following steps are used to identify and delete spurious vectors. Firstly, the vectors are filtered using a local median test. The vectors are rejected and replaced with a genuine secondary peak when the difference between the current velocity vector and the local median velocity of nearby vectors exceeds a predefined tolerance (two times the local median value). Secondly, a vector is valid when the signal-to-noise ratio (SNR) is greater than 1.5. A range filter is used to reject the vector if it exceeds tolerable limits in the axial and radial displacements. A recursive filling process using the local mean is used to fill the holes in the vector field. It starts by filling the holes with the most valid neighbors, then moves on to the ones with the second most valid neighbors. The statistical stationary analysis is conducted for the swirling airflow downstream of the air nozzle for different parameters, such as the laser energy intensity, the number of frames considered for averaging, and repetition rate of the laser in figure \ref{figS2} (in Appendix). It can be observed that the flow field is nearly identical for different parameters, and we also checked that the findings do not alter while using higher values than the parameters considered here. In view of these results, we use 40\% laser energy intensity, 1000 frames for averaging, and 700 Hz repetition rate for the rest of the experiments conducted in this study.

\begin{table}
\begin{center}
\begin{tabular}{cccccc}
Case ~~~~&        $Sw$  ~~~~   & $We$      ~~~~   & ${\bar x_d} = x_d/D_o$           ~~~~  & ${\bar y_d} =y_d/D_o$      ~~~~      & ${\bar z_d} =z_d/D_o$         ~~~~     \\ \hline 
1 ~~~~&     0 ~~~~& 11.40  to 19.47  ~~~~      & 0.01         ~~~~ & 0.89    ~~~~      & 0 ~~~~\\ 
 2 ~~~~&    0.47    ~~~~ & 10.71 to 22.06    ~~~~     & 0.01   ~~~~      & 0.89      ~~~~    & -0.63 to 0.63 ~~~~\\ 
 3  ~~~~&    0.82  ~~~~   & 7.72 to 22.50 ~~~~& 0.01 to 0.51 ~~~~& 0.89 to 4.19 ~~~~& -0.63 to 0.63 ~~~~\\
\end{tabular}
\end{center}
\caption{The dimensionless parameters and location of the dispensing needle normalised with the outer diameter of the swirler ($D_o=20$ mm). } \label{T2} 
\end{table}

\section{Results and discussion}
\label{sec:dis}

\subsection{Breakup morphology for different swirl strengths}\label{sec:mode}

\begin{figure}
\centering
\includegraphics[width=1\textwidth]{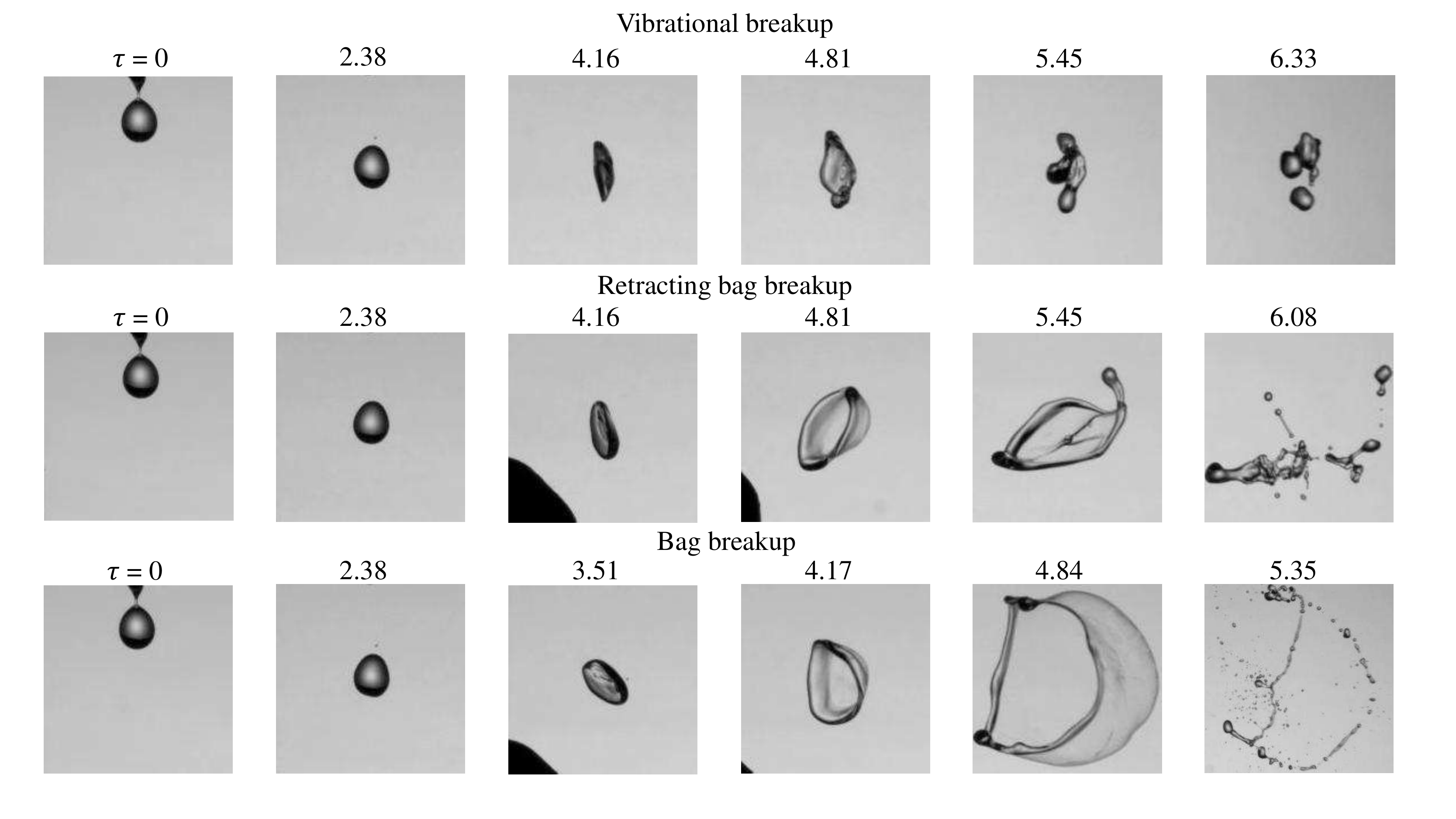}
\caption{Temporal evolution of the breakup dynamics for different values of Swirl number, $Sw$: Top row: $Sw= 0$ (no swirl), middle row: $Sw= 0.47$ (low swirl strength) and bottom row: $Sw= 0.82$ (high swirl strength). In the no swirl case, ${\bar z}_d=0$, but in the swirl cases, ${\bar z}_d = -0.13$. The rest of the parameters are fixed at ${\bar x}_d = 0.01$, ${\bar y}_d = 0.89$ and $We = 11.40$. The number at the top of each panel represents dimensionless time, $\tau=t/T$, where $t$ and $T=d_0 \sqrt{(\rho_l/ \rho_a)}/U$ denote the physical time and the time scale used in our study, respectively. The instant the droplet dispenses from the needle is represented by $\tau = 0$. An animation showing the vibrational (first row), retracting bag breakup (second row) and bag breakup (third row) has been provided as Supplementary Movie 1.}
\label{fig3}
\end{figure}

The primary objective of our study is to explore the interaction of a swirling airflow with a freely falling droplet in an orthogonal configuration, where the droplet can encounter oppose-flow, cross-flow, and co-flow situations in an unsteady manner, resulting in droplet morphologies that are distinct from those observed in the no-swirl configuration. As shown in figure \ref{fig1}, the air flows in the positive $x$ direction, while the droplet falls from the dispensing needle in the negative $y$ direction under the action of gravity. Figure \ref{fig3} depicts the temporal evolution of morphology of an ethanol droplet in no-swirl ($Sw=0$), low swirl ($Sw=0.47$), and high swirl ($Sw=0.82$) conditions. As discussed in the experimental section, the Swirl number $(Sw)$ is varied by changing the vane angle of the swirler. To enable comparison of the present result with the earlier studies, the droplet is ejected at ${\bar z}_d = 0$ (center plane) in the no-swirl condition. But in swirl flow situations, the droplet is ejected at ${\bar z}_d = -0.13$ (negative), thus the droplet interacts with the swirl flow in oppose/cross-flow condition. The rest of the parameters are fixed at $We = 11.40$, ${\bar x}_d = 0.01$ and ${\bar y}_d = 0.89$. The results are presented at different dimensionless time, $\tau=t/T$, where $t$ and $T=d_0 \sqrt{(\rho_l/ \rho_a)}/U$ denote physical time and the time scale used in our study, respectively, and $\tau = 0$ represents the instant when the droplet dispenses from the needle.

For $Sw=0$ (shown in the first row of figure \ref{fig3}), the droplet exhibits a vibrational breakup mode. It can be seen that the droplet enters the aerodynamic field ($\tau = 2.38$) and deforms into a disk shape ($\tau=4.16$). In this case, the aerodynamic force is insufficient to overcome the viscous and surface tension forces, causing shape oscillations in the droplet (see, $\tau=4.16$ to $5.45$). However, due to capillary instability, the droplet breaks into a few smaller droplets of comparable size at a later time. The dynamics observed for this set of parameters is qualitatively similar to that reported in the earlier studies (e.g. \cite{Jain2015,kulkarni2014bag}).

A new type of breakup mode (termed as `retracting bag breakup') is observed for an intermediate swirl strength ($Sw=0.47$) as shown in the second row of figure \ref{fig3}. In this case, when the droplet enters the swirl flow airstream, the lower part of the droplet first enters the swirling flow region, and the shape of the droplets is quickly converted into a slightly tilted disk shape from a near-spherical shape as the droplet encounters an oppose-flow condition (see, at $\tau=4.16$ in figure \ref{fig3}). The aerodynamic field has a high velocity in the shear zone, a low velocity in the wake of the vanes, and a negative velocity in the centre region due to the recirculation created by the imposed swirl flow (refer to \S\ref{sec:flow}). The disk shape that faces the strong swirl field bulges and creates a bag in the shear zone, while the bottom portion of the disk shape remains thicker as it encounters the low-velocity zone. The droplet exhibits the opposite condition when it first enters the vane's wake and migrates towards the shear layer. As a result, liquid accumulated in the wake region and formed a large node at the beginning of the bag breakup process ($\tau= 4.81$). Subsequently, under the influence of the aerodynamic force, the partial disk shape participates in the bag breakup process. As lower part of the droplet comes in wake of the vane, the velocity at the upper side of the disk becomes high, while the velocity at the lower side of the disk is low, causing the disk to rotate clockwise in the $x-y$ plane ($\tau=4.16 - 6.08$, see Figure \ref{figS4} in Appendix). As the part of the droplet passes through the wake region, negative pressure drives the bag film in the direction opposite to the bag growth. Thus, while the bag retracts in the upper half of the drop, it remains intact in the lower half ($\tau=5.45$). The instability is induced on the drop surface in such a way that the nodes in the upper portion of the droplet move quicker while the nodes in the lower portion of the droplets move slower. Therefore, the ring is stretched and the bag is retracted by the airflow continuously. Further retraction of the bag ($\tau=6.08$) leads to the development of capillary instability, which causes the bag to burst in opposite direction and the rim to disintegrate, which is distinct from the conventional bag breakup process. At later times, as the liquid enters the negative velocity region generated by the recirculation zone, the lower part of the liquid disintegrates from the larger droplet. The retraction bag bursting process can be understood as the consequence of a change in the aerodynamic field, which is different from that in the no swirl case. An animation showing the vibrational, retracting bag breakup and bag breakup has been provided as Supplementary Movie 1.

As the Swirl number increases from $Sw=0.47$ to 0.82, the mode changes from retracting bag breakup to bag breakup mode (third row in figure \ref{fig3}). The dynamics observed at the early times is similar to that of the low swirl case ($Sw=0.47$). The drop deforms into a tilted disk shape ($\tau=3.51$) as it enters the swirl flow region in contrast to the straight disk confronting the airstream in the no swirl case as shown in the first row of figure \ref{fig3} (also see, \cite{guildenbecher2009secondary, rimbert2020spheroidal, jackiw2021aerodynamic}). In the swirl flow, the orientation of the disk depends on the strength of the swirl and its interaction with the swirl flow stream (refer to the discussion in \S\ref{sec:flow}). It can be seen that the deformed disk is slanted more for $Sw=0.82$ than for $Sw=0.47$. \cite{soni2020deformation} also reported the deformed disk in various orientations when a drop interacts with a uniform airstream at oblique angles. In this case, the disk was unable to cross the shear layer region. At $\tau=4.17$, the disk orients itself vertically enabling the entire disk to participate in the bag breakup process. Thus efficient energy transfer occurs and the bag expands in the direction of the swirling airflow resulting in a thinner rim as compared to the low swirl case. The formation of multiple nodes is caused by the multi-directional stretching of the droplet, which is higher than that in the no-swirl and low-swirl cases. The formation of multiple nodes due to the swirling flow causes the faster stretching of the bag and rim ($\tau=4.84$). The bag and rim then disintegrate into tiny droplets ($\tau=5.35$). In this case, the droplet breakup is mostly caused by high shear velocity in the shear region when the swirl strength is strong. 

\begin{figure}
\centering
\includegraphics[width=0.95\textwidth]{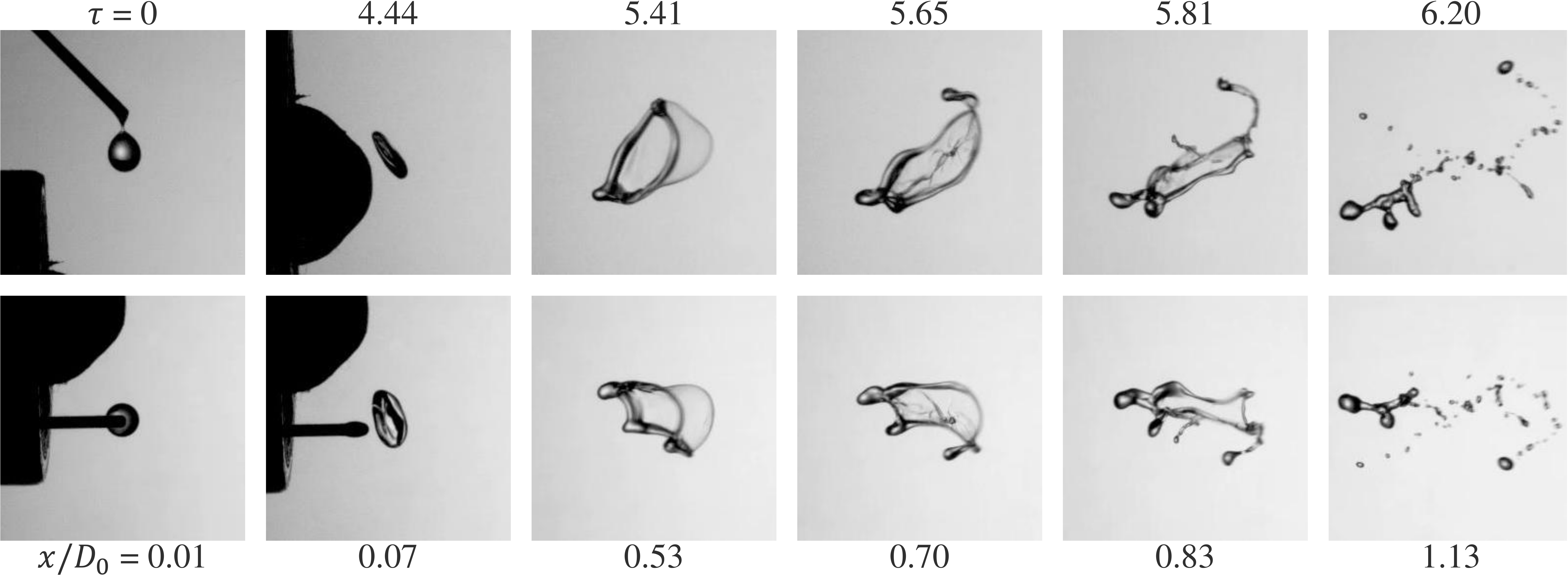}
\caption{Temporal evolution of the retracting bag breakup morphology for $We = 7.72$ and $Sw = 0.82$ in two views. The front $(x-y)$ and top $(x-z)$ views are shown in the top and bottom rows, respectively. Here, $\bar x_d = 0.01$, $\bar y_d = 0.89$, and $\bar z_d = -0.48$. The dimensionless time ($\tau$) and the corresponding streamwise location of the drop are mentioned at the top and bottom, respectively. An animation showing the front and top views has been provided as Supplementary Movie 2.}
\label{fig_RBB}
\end{figure}

The retracting bag breakup mode is apparent only when the droplet enters the vane's wake and the central recirculation zone, which can be explained as follows. Two important factors influencing the breakup process are the stretching rate and residence time of the drop in the high airstream zone. \cite{soni2020deformation} reported that the obliquity of the airstream influences the stretching factor, which in turn influences the breakup phenomenon (also, refer to the discussion in \S\ref{subsec:evolution}). As the angle of obliquity of the airstream increases, the droplet's rectilinear trajectory changes into a curvilinear trajectory. When a droplet is exposed to swirl air, it experiences a larger stretching factor than when it is exposed to no swirl air. Moreover, the droplet encounters a radial centrifugal force in swirl flow, which prevents the droplet from penetrating the airstream. This force increases as the swirl strength increases. Thus, the droplet easily penetrates the shear region in the low swirl case ($Sw=0.47$) and experiences a low velocity in the wake of the vanes and a forward velocity in the shear zone. Thus, the droplet develops the retracting bag breakup mode. On the other hand, in the case of strong swirl flow ($Sw=0.82$), the drop does not fully reach the wake region created by the vanes and the recirculation zone created by the swirl flow, thus, resulting in a regular bag breakup mode. 

The droplet breakup phenomenon for another set of parameters ($\bar x_d = 0.01$, $\bar y_d = 0.89$, $\bar z_d = -0.48$, $We = 7.72$ and $Sw = 0.82$) is presented in figure \ref{fig_RBB} (the corresponding animation is provided as Supplementary Movie 2). Here, both the front and top views capture using the two high-speed cameras are depicted. In this case, the disc easily enters and passes through the shear region as the aerodynamic force is low. The droplet exhibits the retracting bag breakup mode for this set of parameters as well, indicating that it is dependent on the Swirl and Weber numbers, and the location of the dispensing needle (also see, figure \ref{figS4} in Appendix). This point is further elaborated in \S \ref{sec:map} where we present the regime map that demarcates different breakup modes for various parameters.

\subsection{Regime map} \label{sec:map}

\begin{figure}
\centering
\includegraphics[width=0.8\textwidth]{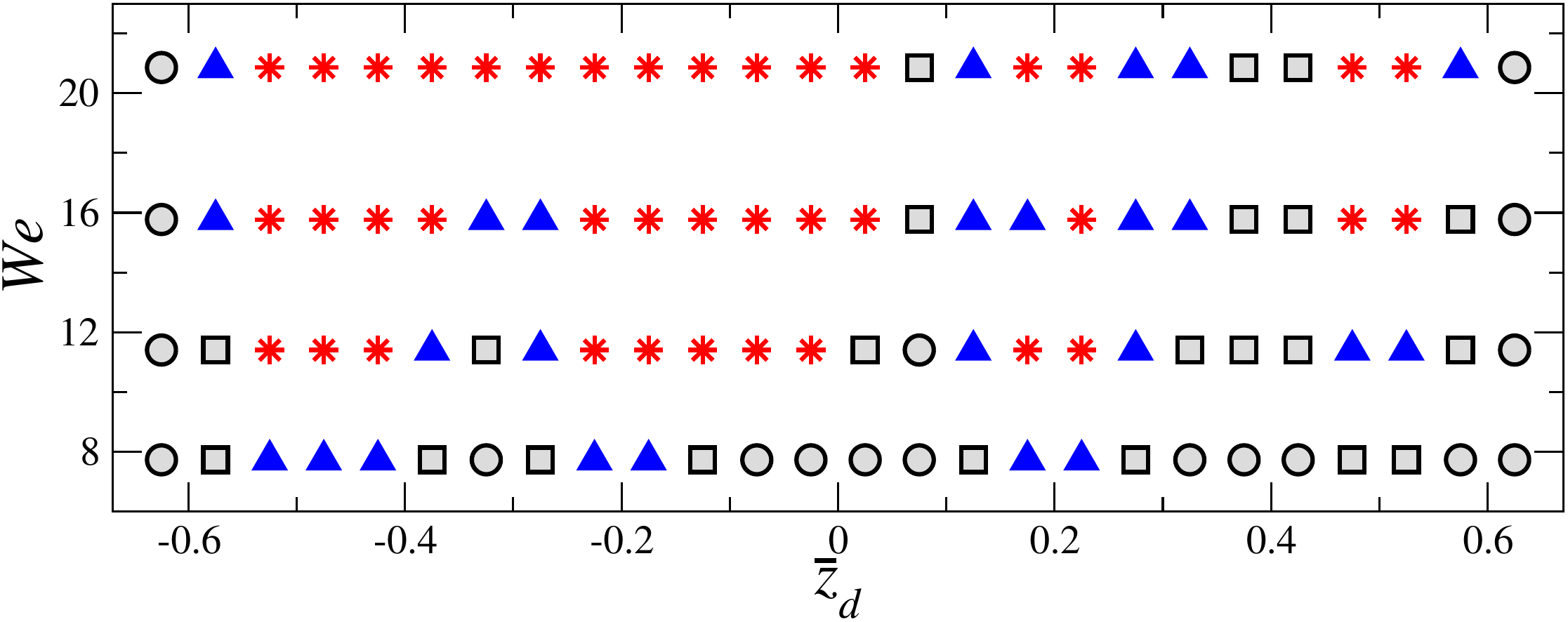}
\caption{Regime map demarcating no-breakup (circle), vibrational breakup (square), retracting bag breakup (triangle) and bag breakup modes (star) in the $\bar z_d - We$ space. The rest of the parameters are fixed at $\bar x_d = 0.01$, $\bar y_d = 0.89$ and $Sw = 0.82$.}
\label{fig:6}
\end{figure}

Figure \ref{fig:6} depicts a regime map demarcating different modes, namely no-breakup (circle), vibrational breakup (square), retracting bag breakup (triangle) and bag breakup modes (star), in the $\bar z_d - We$ space. The rest of the parameters are fixed at $\bar x_d = 0.01$, $\bar y_d = 0.89$ and $Sw = 0.82$. Experiments are repeated five times for each set of parameters to ensure repeatability. The droplet encounters the airstream in oppose, cross, and co-flow conditions as we move the needle from negative to positive values of $\bar z_d$ (see, figure \ref{figS3} and the associated discussion in \S\ref{sec:flow}). For $\bar z_d=-0.63$ and 0.63, as the droplet falls under the action of gravity, it does not interact with the swirl flow and remains nearly spherical with minor shape oscillations but without breakup (circle symbols), indicating that these locations are outside the swirl airflow region. This implies that the aerodynamic potential and stretching rate are insignificant enough to disintegrate even the initial droplet. It can be observed that as we move the dispensing needle from $\bar z_d=-0.63$ to 0.63, the droplet only exhibits no breakup, vibrational and retracting bag breakup modes for $We=7.72$ (low airstream velocity). This is owing to the resultant lower aerodynamic force and low stretching rate in these cases. Close inspection also reveals that there are three low velocity zones created by the swirler's vanes near $\bar z_d \approx -0.33$, $0.08$ and $0.40$. For $We=11.8$ (increasing the velocity of the airstream), we observe bag breakup mode in addition to the other modes. The retracting bag breakup modes appear at the edges of low velocity zones due to differential velocity interaction. The droplet shows bag breakup mode for negative $\bar z_d$ locations (oppose-flow and cross-flow conditions) and positive $\bar z_d$ locations slightly away from the centreline (co-flow condition).  A further increase in the Weber number creates a more favorable condition for the bag breakup, especially for negative $\bar z_d$. The following is an explanation for the behavior. In opposed flow conditions, efficient aerodynamic energy transfer occurs due to the long droplet-air interaction time; however, in co-flow conditions, the local migration of the droplet and the airflow are in the same direction, resulting in a short residence time for efficient energy transfer.

\begin{figure}
\centering
\vspace{1mm}
\includegraphics[width=0.8\textwidth]{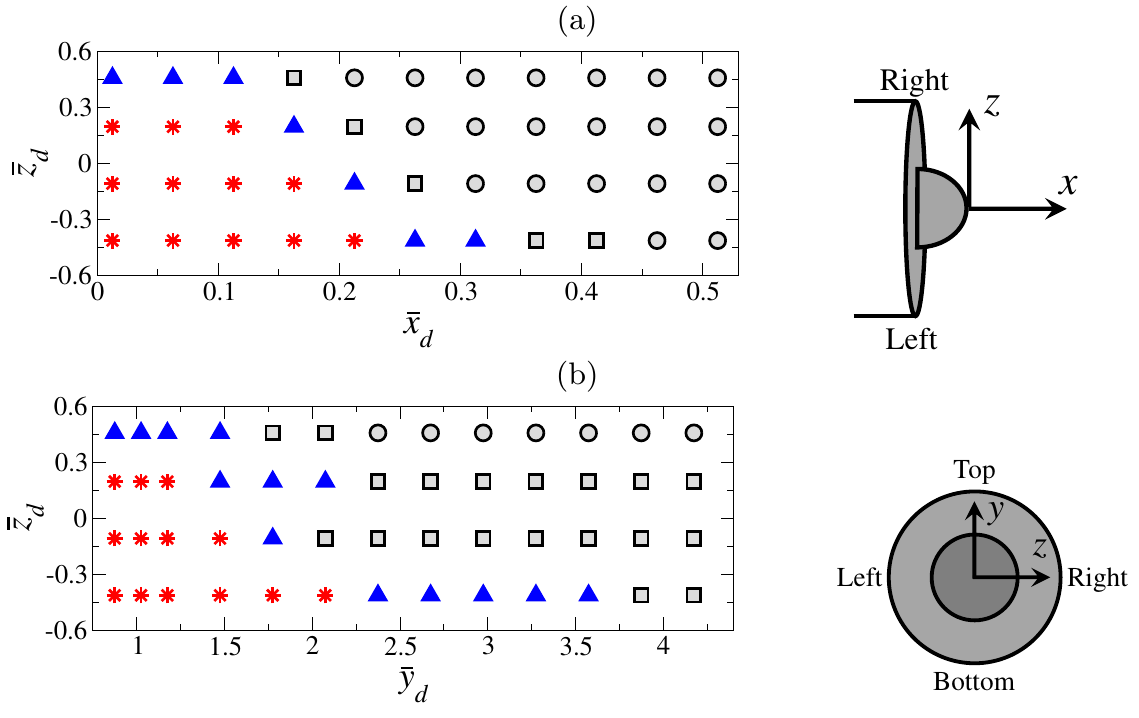} \\
\caption{Regime map showing the no-breakup (circle), vibrational breakup (square), retracting bag breakup (triangle) and bag breakup (star) modes in (a) $\bar x_d-\bar z_d$ plane at $\bar y_d = 0.89$, and (b) $\bar y_d-\bar z_d$ plane at $\bar x_d = 0.01$. The right side panels show the directions with respect to the nozzle position in the $x-z$ plane (top view) and $y-z$ plane (side view). The rest of the parameters are $Sw = 0.82$ and $We=11.40$.}
\label{fig:7}
\end{figure}

It is observed that the swirler vanes create three dead zones, and four high airstream flow zones in between the dead zones and outer no breakup zones. To explore the influence of varying the coordinate of the dispensing needle on droplet breakup morphology, the high airstream zones at $\bar z_d=-0.48$, $-0.13$, $0.23$ and $0.53$ are considered. Figure \ref{fig:7}(a) and (b) show the various breakup modes observed by changing the location of the dispensing needle in the $\bar x_d-\bar z_d$ plane at $\bar y_d = 0.89$, and $\bar y_d-\bar z_d$ plane at $\bar x_d = 0.01$, respectively. Here, $Sw = 0.82$ and $We=11.40$. It can be seen in figure \ref{fig:7}(a) that there is a transition from the bag breakup mode to no breakup via the retracting bag breakup and vibrational breakup modes as $\bar x_d$ is increased (moving away from the nozzle in the streamwise direction) for any $\bar z_d$ location. The dilution of the flow field in the downstream direction from the tip of the air nozzle causes the various breakup modes (refer to our discussion in \S\ref{sec:flow}). The droplet experiences an oppose-flow condition at negative $\bar z_d$ due to the helical motion of the swirl airstream, which is a key favorable condition for the bag fragmentation process. Due to the transition from favourable to unfavourable conditions from a bag breakup perspective, the axial extent downstream of the nozzle/swirler for bag breakup mode reduces when the droplet dispensing needle traverses in from $-\bar z_d$ to $\bar z_d$ direction. Near the centerline of the airstream, the droplet experiences a partial cross-flow with an oppose-flow and a co-flow conditions in the negative and positive sides of $\bar z_d=0$, respectively. As a suspended droplet in this zone is promptly ejected by the swirl flow field into the low-velocity region, the probability of the bag breakup is low in the downstream direction. Similarly, a suspended droplet in a peripheral location (see at $\bar z_d = 0.48$) encounters the airstream in a co-flow condition, and with the little interaction time, the deformed droplet goes out of the main airstream. Therefore, the bag breakup probability is less for the given operating parameters.

\begin{figure}
\centering
\hspace{0.0 cm} (a)   \hspace{5.5cm}  (b) \\
\includegraphics[width=0.4\textwidth]{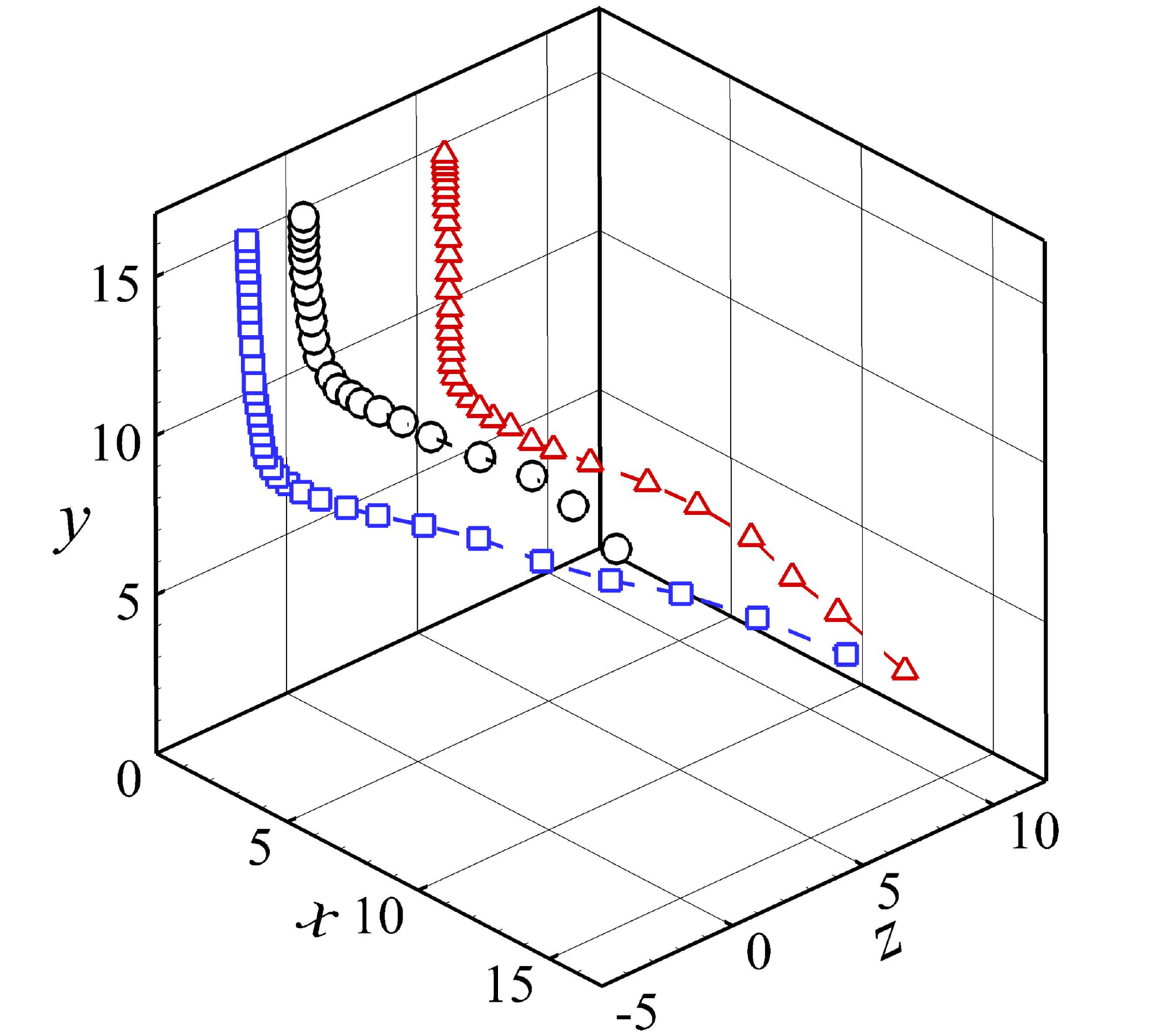} \hspace{5mm} \includegraphics[width=0.4\textwidth]{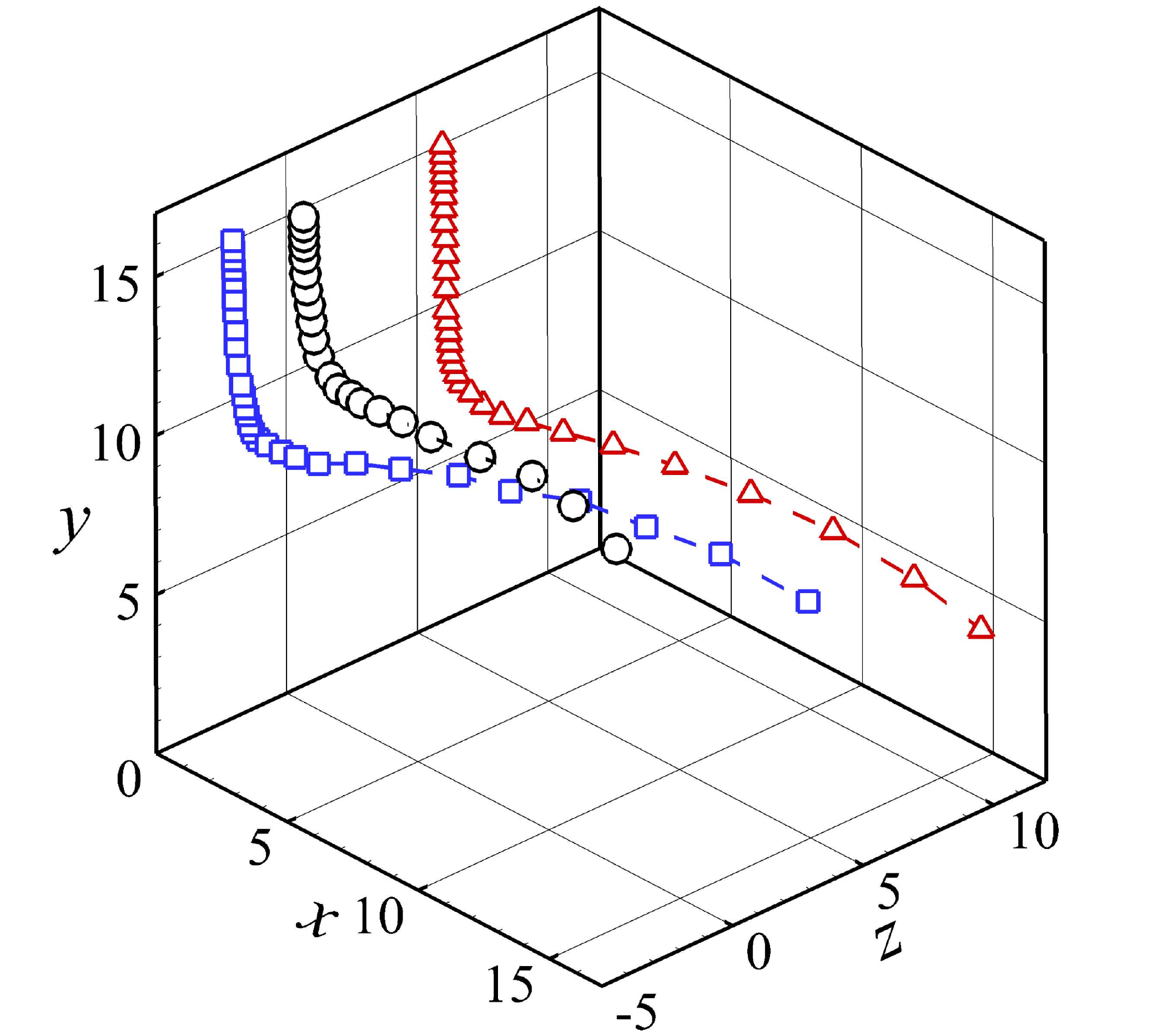} \\ 
\hspace{0.0 cm} (c)   \hspace{5.5cm}  (d) \\
\includegraphics[width=0.4\textwidth]{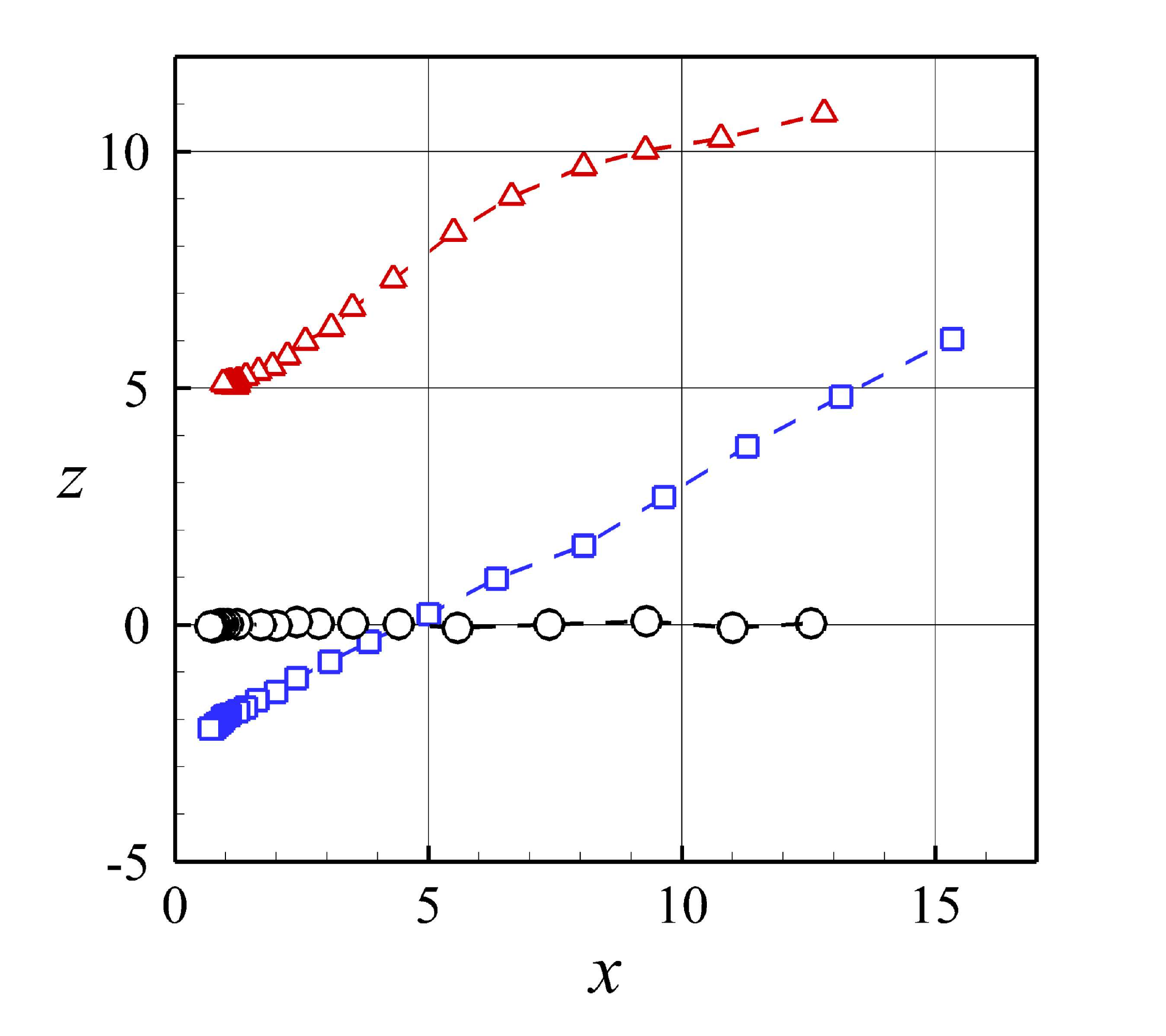} \hspace{5mm} \includegraphics[width=0.4\textwidth]{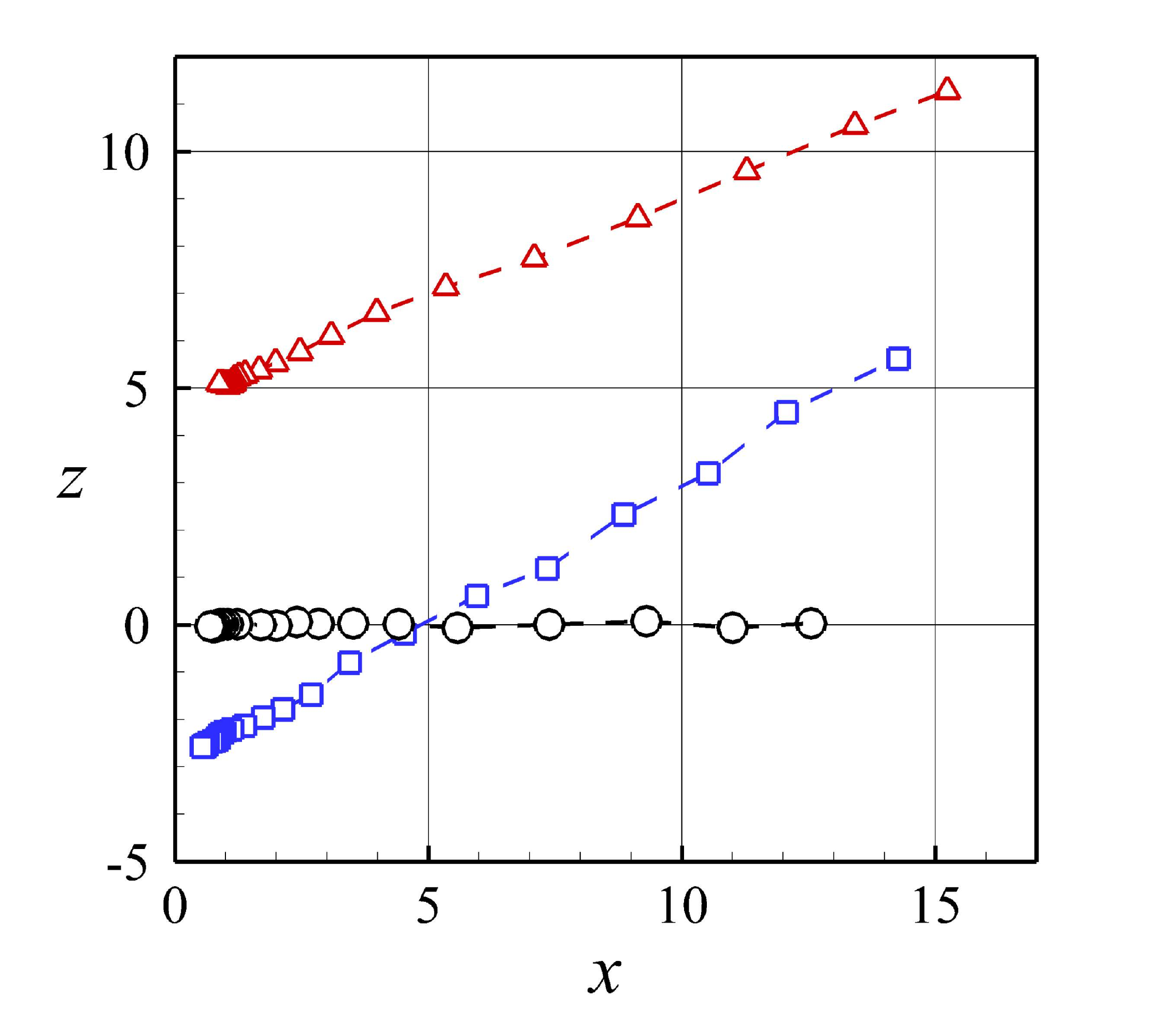}\\
\hspace{0.0 cm} (e)   \hspace{5.5cm}  (f) \\
\includegraphics[width=0.4\textwidth]{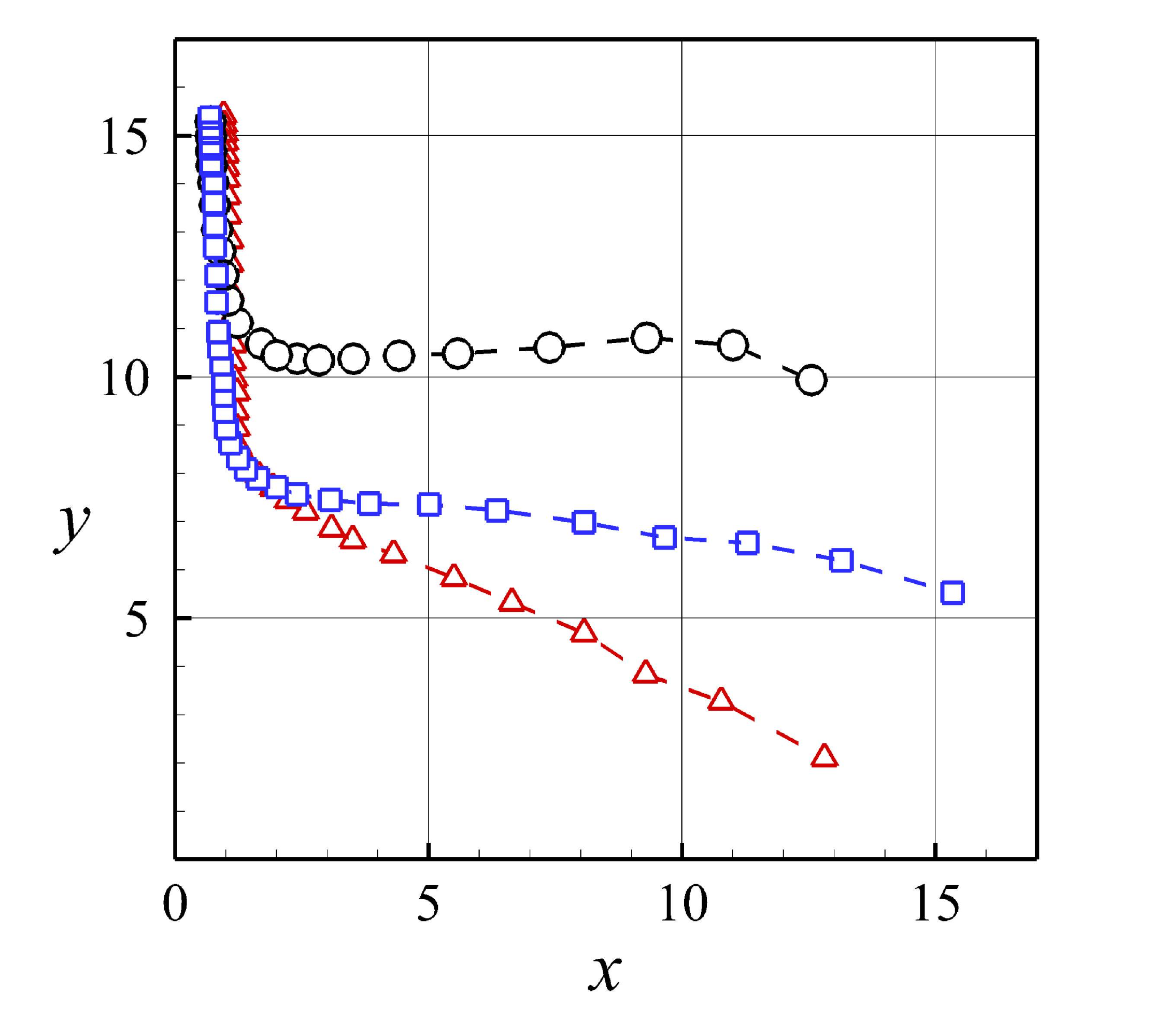} \hspace{5mm} \includegraphics[width=0.4\textwidth]{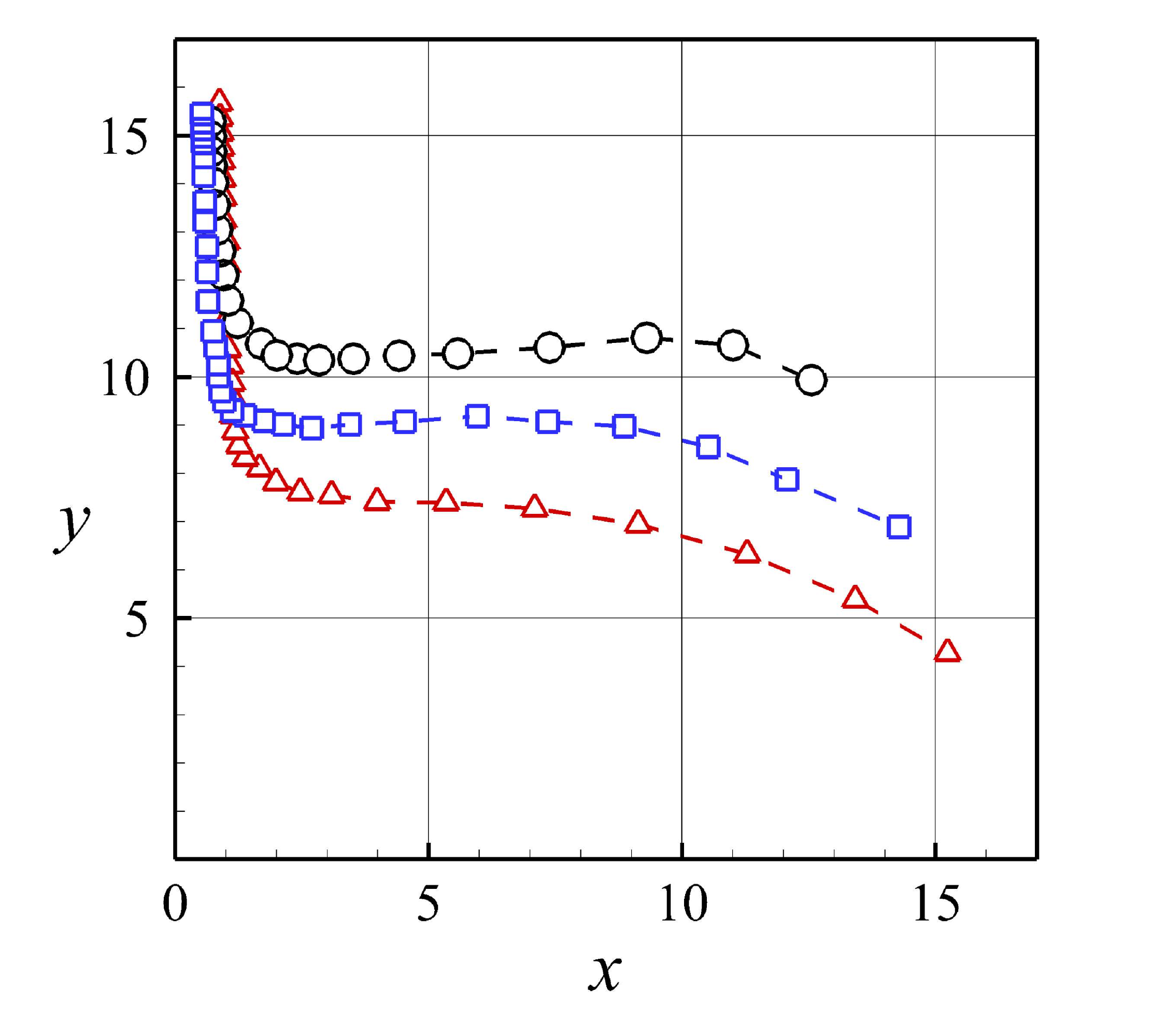} 
\caption{Trajectories of the droplet for $We=11.40$ in different conditions. Panels (a,c,e): $Sw=0.47$; and (b,d,f): $Sw=0.82$. The three dimensional view of the droplet's trajectories is shown in panels (a,b). The projections on the $x-z$ and $x-y$ planes are plotted in panels (c,d) and (e,f), respectively. The results represented by blue squares and red triangles in all panels are associated with $\bar z_d = -0.13$ and 0.23, respectively. The no swirl case ($Sw=0$) with $\bar z_d=0$ is presented (black circles) for comparison purpose. Here, the locations $x, y$, and $z$ are in mm.}
\label{fig:8}
\end{figure}

The height of the dispensing needle, $\bar y_d$ can also play an important role in the breakup dynamics. Thus, we investigate the droplet breakup modes in the $\bar y_d-\bar z_d$ plane at $\bar x_d = 0.01$ (figure \ref{fig:7}(b)). The inertial kinetic energy of the droplet migrating in the downward direction is related to the height of the dispensing needle, which increases as the height of the dispensing needle increases. A droplet with a high downward velocity (injected at a high $\bar y_d$ location) enters and travels quickly through the shear layer zone without being influenced by the swirl airstream. Thus, at a fixed negative $\bar z_d$, i.e. in an oppose-flow and cross-flow conditions (see, for instance, $\bar z_d=-0.48$ and $-0.13$), increasing the droplet dispensing height changes from the bag breakup to retracting bag breakup to vibrational breakup to no breakup mode. The transition from bag to retracting bag breakup mode occurs at a lower value of $\bar y_d$ when we move the dispensing needle from $\bar z_d=-0.48$ to $0.23$. This is because the swirler has a dome-like shape at the center that results in an annular flow from the air nozzle. As the height of the dispensing needle is increased (see, for instance, at $\bar z_d=-0.13$ and $\bar y_d=2.09$), the droplet interacts with the upper part of the annular flow and pushes outside and downward. The droplet then enters the central recirculation zone produced by the dome and only exhibits the vibrational breakup mode even for higher values of $\bar y_d$. At $\bar z_d = 0.23$, the droplet experiences a cross-flow with co-flow condition. The upper annular flow pushes the droplet towards the annular flow on the right side. In the co-flow configuration ($\bar z_d=0.53$), on the other hand, the bag breakup mode is not observed as it interacts with the airstream at the right side annular flow region. At lower heights, the potential energy of the droplet is low, and hence the droplet has been influenced by the flow and tries to be in the potential core so that the droplet exhibits bag breakup mode. At moderate $\bar y_d$ locations, we have observed the retracting bag breakup. In contrast, for higher heights, due to the associated higher kinetic energy, the droplet easily penetrates in the potential core, resulting in no breakup or vibrational breakup mode. From the above discussion, one can conclude that drop breakup morphology is strongly influenced by the several key factors, such as aerodynamic field, initial momentum and residence time of the droplet.

\subsection{Droplet trajectory} \label{sec:traj}

In order to gain a better understanding of the interaction of the droplet with the swirl airstream, in figure \ref{fig:8}, we plot its trajectory for various conditions, such as different values of the Swirl number and location of the dispensing needle. The Weber number is fixed at $We = 11.40$. The trajectory of the droplet injected at $\bar z_d = 0$ for the no swirl case ($Sw = 0$) is shown (black circles) for the reference displaying the vibrational breakup mode. In this case, the droplet enters the airstream but does not penetrate the potential core zone; instead, it deflects away from the shear layer region. The conditions experienced by the droplet for this set of parameters favour the vibration breakup mode. The high ($Sw = 0.82$) and low swirl ($Sw = 0.47$) cases considered in figure \ref{fig:8} exhibit the bag breakup and the retracting bag breakup modes, respectively. As the swirler rotates in the clockwise direction, the droplet migrates from the negative $z$ to the positive $z$ direction (figure \ref{fig:8}(c) and (d)). The droplet experiences an upward force due to swirling flow when $\bar z_d$ is negative ($\bar z_d = -0.13$) and thus, resulting in a higher curvilinear trajectory in the high swirl case (figure \ref{fig:8}(b), (d) and (f)) as compared to the low and no swirl cases (figure \ref{fig:8}(a), (c) and (e)). Inspection of figure \ref{fig:8}(e) and (f) also reveals that in contrast to the no swirl case where the droplet follows mostly a rectilinear trajectory, in the swirl case ($Sw=0.47$), the droplet undergoes a curvilinear trajectory and approaches the central recirculation region through the dead zones of the vanes, which leads to retracting bag breakup mode for $\bar z_d = -0.13$. This curvilinearity in the trajectory is enhanced as we increase the Swirl number. It can be seen that the droplet travels a compact path length as it undergoes a bigger turning in the swirl cases compared to the no swirl condition. By comparing the left side and right side panels of figure \ref{fig:8}, we observe that the flow configuration (oppose-flow: $\bar z_d = -0.13$ and co-flow: $\bar z_d = 0.23$) has a significant effect on the droplet trajectory due to the change in the aerodynamic field for positive and negative values of $\bar z_d$. An oppose-flow condition limits the droplet penetration to the potential core region of the main airstream. Thus, the drop experiences a retarding velocity, resulting in a smaller drop penetration in the bag formation process than in the co-flow situation (see, the trajectories marked by rectangles and triangles in figure \ref{fig:8}).

\subsection{Analysis of the imposed swirl flow using stereo-PIV} \label{sec:flow}

\begin{figure}
\centering
\includegraphics[width=0.95\textwidth]{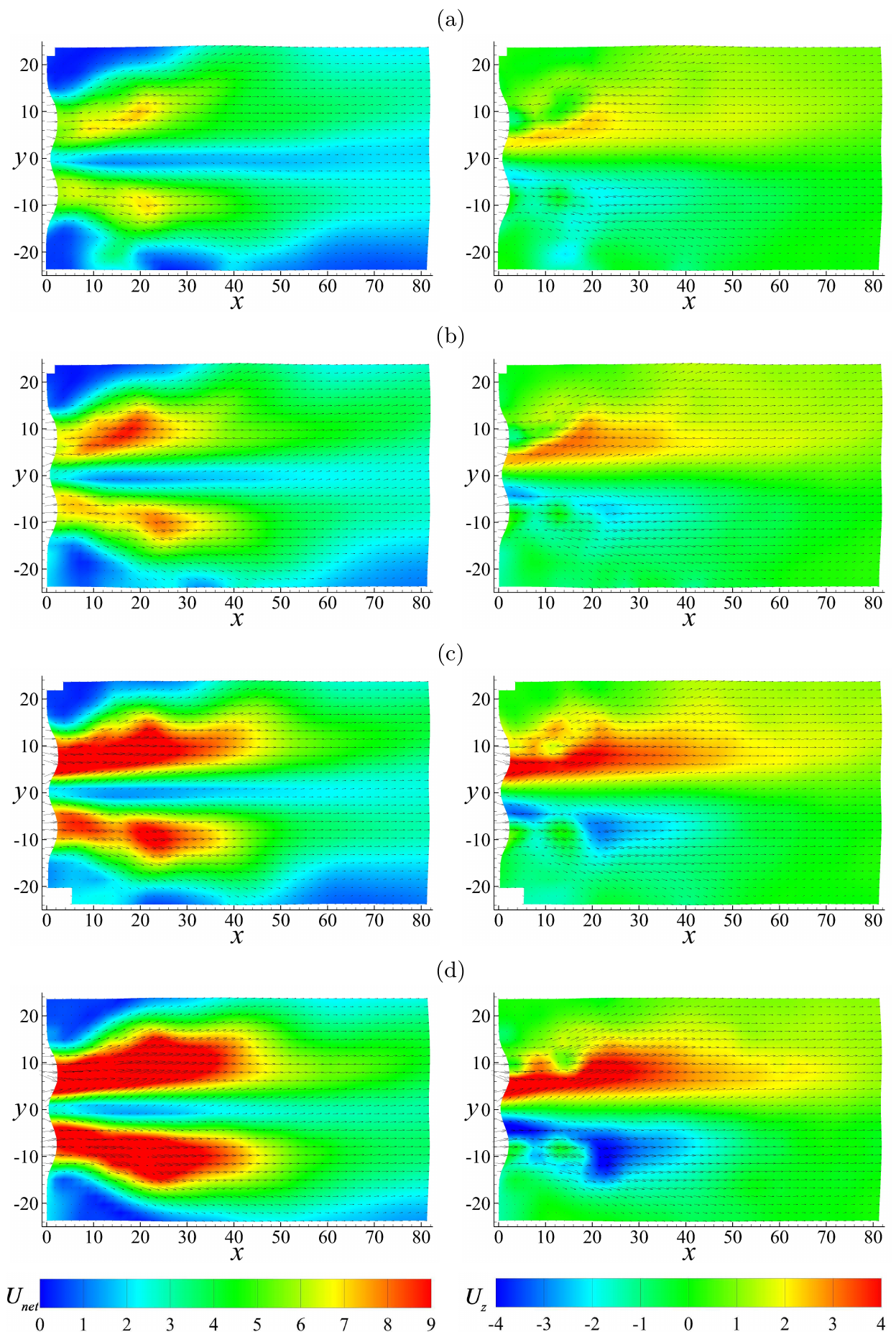}  \\
\caption{The contours of the resultant velocity, $U_{net} (= \sqrt{U_x^2+U_y^2+U_z^2})$ (left panels) and $U_z$ (right panels) in the $x-y$ plane at ${\bar z}_d = 0$  for different Weber numbers: (a) $We = 7.72$, (b) $We = 11.40$, (c) $We = 15.77$ and (d) $We = 20.86$. In all the cases, $Sw = 0.82$. The $x$ and $y$ labels are in mm, and $U_{net}$ and $U_z$ are in m/s.}
\label{fig:2}
\end{figure}

\begin{figure}
\centering
\includegraphics[width=0.95\textwidth]{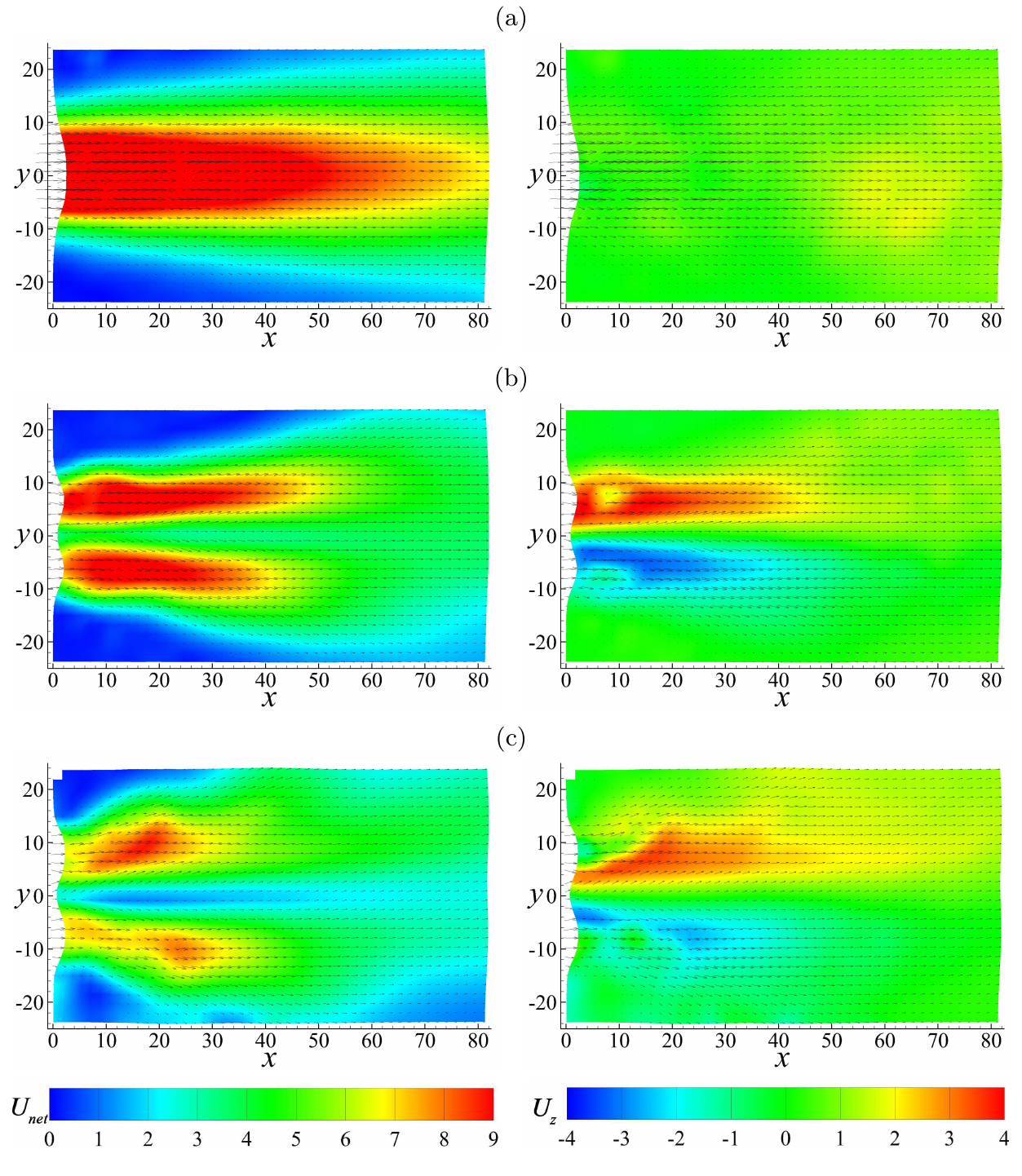} \\ 
\caption{The contours of the resultant velocity, $U_{net} (= \sqrt{U_x^2+U_y^2+U_z^2})$ (left panels) and $U_z$ (right panels) in the $x-y$ plane at ${\bar z}_d = 0$ for different Swirl numbers: (a) $Sw = 0$, (b) $Sw = 0.47$ and (c) $Sw = 0.82$. In all the cases, $We = 11.40$. The $x$ and $y$ labels are in mm, and $U_{net}$ and $U_z$ are in m/s.}
\label{fig:3}
\end{figure}

\begin{figure}
\centering
\includegraphics[width=0.95\textwidth]{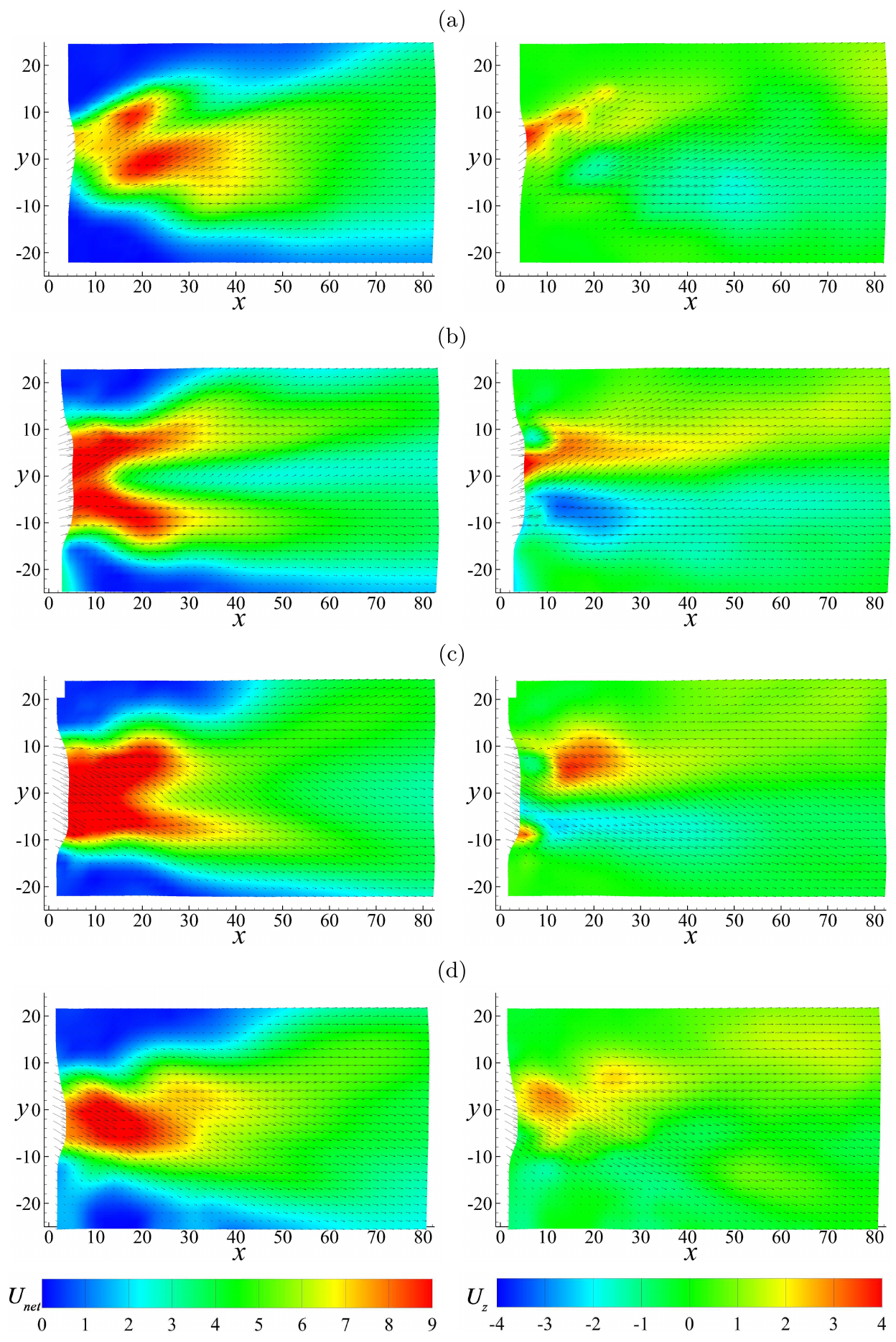}  \\
\caption{The contours of the resultant velocity, $U_{net} (= \sqrt{U_x^2+U_y^2+U_z^2})$ (left panels) and $U_z$ (right panels) in the $x-y$ plane at (a) $\bar z_d = -0.48$, (b) $\bar z_d = -0.13$, (c) $\bar z_d = 0.23$ and (d) $\bar z_d = 0.53$. In all the cases, $Sw = 0.82$ and $We = 11.40$. The $x$ and $y$ labels are in mm, and $U_{net}$ and $U_z$ are in m/s.}
\label{figS3}
\end{figure}

As the imposed swirl flow field essentially influences the breakup phenomenon of the drop, we analyze the flow field for different swirl strengths (characterized by the Swirl number), airflow rates in the nozzle (characterized by the Weber number), and the location of the droplet migration plane using a stereo-PIV set-up as shown in figure \ref{figS1} in Appendix. Figures \ref{fig:2} and \ref{fig:3} depict the contours of the resultant velocity, $U_{net} (= \sqrt{U_x^2+U_y^2+U_z^2})$ and $U_z$ overlapped with the velocity vectors for different Weber and Swirl numbers, respectively. Here, $U_x$, $U_y$ and $U_z$ denote the components of the velocity vector in the $x$, $y$, and $z$ directions, respectively. It can be seen that the resulting flow is an annular flow. In figure \ref{fig:2}, for a fixed value of the Swirl number, we observe that increasing the Weber number increases the magnitude of velocity, leading to stronger annular flow with a recirculating zone in the centerline region. Thus, a droplet encounters a more aerodynamic force as the Weber number increases. The contours of $U_z$, which indicate the flow rotating clockwise, demonstrate the droplet movement from left to right when dispensing in the flow regime. The wake regions due to the vanes are also clearly evident in the flow field showing $U_z$ contours. Similarly, in figure \ref{fig:3}, we present the contours of $U_{net}$ and $U_z$ for different Swirl numbers for a fixed value of the Weber number ($We = 11.40$). It can be observed that while a symmetrical potential core region is apparent in the no swirl case (figure \ref{fig:3}(a)), in the swirl cases (figure \ref{fig:3}(b) and (c)), an annular flow due to the dome shape at the center of swirler develops in the downstream, that dramatically alter the breakup mode as discussed in \S\ref{sec:mode}. The radial component of the velocity vector is negligible in the absence of swirl, but it becomes substantial in the presence of swirl. The outward radial component of the velocity is enhanced while the axial velocity component is decreased as the strength of swirl flow (i.e. the Swirl number) is increased (figure \ref{fig:3}(b) and (c)). The characteristics of the swirl flow (contours of $U_{net}$ and $U_z$) in the $x-y$ plane taken at different $\bar z$ locations (namely, $\bar z_d=-0.48$, $-0.13$, $0.23$ and $0.53$) are presented in figure \ref{figS3} for $We=11.40$ and $Sw=0.82$. Figure \ref{figS3}(a) (for $\bar z_d=-0.48$) corresponds to the $x-y$ plane at the extreme left side of the swirler. In this case, the influence of the vane as an obstacle to the flow is clearly seen at $x \approx 16$mm and $y \approx 2$ mm. The contours of $U_z$ for $\bar z_d=-0.48$ indicate that the droplet experiences an oppose-flow condition (see the upward moving velocity vectors in figure \ref{figS3}(a) and (b). As we move towards the central location (at $\bar z_d=-0.13$), the droplet still experiences an oppose-flow condition, albeit of lower strength as compared to $\bar z_d=-0.48$. On the other hand, when $\bar z_d$ is positive ($\bar z_d=0.23$ and 0.53), i.e. in the right side of the swirler, the droplet experiences a co-flow configuration (as indicated by the downward velocity vectors in figure \ref{figS3}(c) and (d). As noted in the introduction, a few researchers have previously used high-speed photography and particle image velocimetry (PIV) techniques to explore the features of swirl flow from a nozzle, albeit without droplets \citep{merkle2003effect,rajamanickam2017dynamics,kumar2019large,patil2021air,soni2021liquid}). They also reported a similar flow field.

%\clearpage
\subsection{Theoretical modeling: Rayleigh-Taylor (RT) instability}
\label{subsec:theory}

The Rayleigh-Taylor (RT) instability occurs when a lighter fluid penetrates into a heavier fluid \citep{sharp1983overview} which plays an important role in the secondary breakup process of a droplet interacting with the airstream \citep{guildenbecher2009secondary}. The Atwood number, $At (\equiv (\rho_{l}-\rho_{a})/(\rho_{l}+\rho_{a}))= 0.997$, characterises the density contrast in our case. Thus, the interface of the droplet (heavier fluid) becomes unstable due to the penetration of the air phase into the liquid (ethanol) phase and perturbations (undulations) of a certain wavelength form at the droplet interface. These surface perturbations grow exponentially with time due to energy transfer from the airstream. Subsequently, the droplet undergoes fragmentation due to these instabilities. Several researchers have conducted linear stability analyses \citep{sahu2009linear} and direct numerical simulations \citep{sahu2009pressure,sahu2011multiphase} to study the effects of viscosity contrast, density contrast and surface tension on the development of RT instabilities in shear flows. The RT instability is also commonly known as fingering instability as the small perturbations quickly develop into interpenetrating fingers of the heavier fluid \citep{sharp1983overview}. It was also shown that when the wavelength of the disturbances created by the RT instability is smaller than the drop diameter, the fingers grow further and fragmentation of the drop occurs \citep{joseph1999breakup}. 

\begin{figure}
\centering
\includegraphics[width=0.95\textwidth]{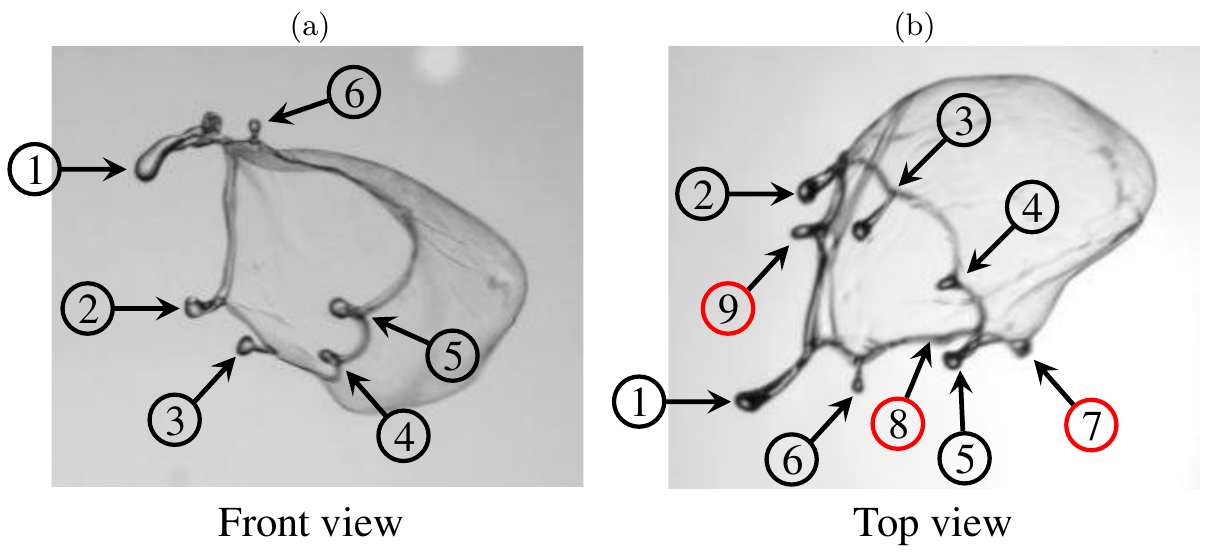}
\caption{Counting the number of fingers using the (a) front view ($x-y$) and (b) top view ($x-z$). The parameters are $We=14.54$, $Sw=0.82$, $\bar x_d = 0.01$, $\bar y_d = 0.89$ and $\bar z_d = -0.13$. While only six fingers (numbered in black circles) are visible in the front view, three more fingers (number in red circles) are apparent in the top view.}
\label{fig:9}
\end{figure}

Figure \ref{fig:9}(a) and (b) illustrates the development of fingers at the onset of a bag burst in two views, namely the front ($x-y$) and top ($x-z$) views, respectively. The parameters are $We=14.54$, $Sw=0.82$, $\bar x_d = 0.01$, $\bar y_d = 0.89$ and $\bar z_d = -0.13$. It can be observed that while just six fingers are visible in the front view, three additional fingers (total nine fingers) are apparent in the top view. In the theoretical modeling (discussed below), as the total number of fingers will be used to compute the maximum wavelength ($\lambda_m$) of the RT instability we use both the views to compute the total number of fingers. It should be noted that previous studies (e.g. \cite{zhao2010morphological}) solely employed the front view to estimate the number of fingers. The maximum wavelength ($\lambda_m$) is given by 
\begin{equation} \label{eq:lambdaexp}
 \lambda_m = \frac{\pi D_m}{N_f},
\end{equation}
where $D_m$ is the maximum disk diameter of the droplet and $N_f$ is the number of fingers counted experimentally at the onset of the breakup, as demonstrated in figure \ref{fig:9}. In liquid jet breakup and droplet breakup under straight airstream, $\lambda_m = \sqrt{3}\lambda_c$, wherein $\lambda_c = {2\pi \sqrt{\sigma_l/\rho_l a}}$ is the critical wavelength of RT instability and $a$ is the acceleration of the drop \citep{marmottant2004spray, varga2003initial, zhao2010morphological, gao2015breakup}. In our study, the front ($x-y$) and top ($x-z$) views acquired using high-speed cameras and processed with the digital image processing NIH IMAGE (ImageJ) software are utilized to determine the acceleration of the drop. From our experiments, in the case of straight airflow $(Sw = 0, \bar z_d = 0)$, we found that $\lambda_m = 4.47$ mm, which is closed to that obtained using the classical analytical expression, $\lambda_m = \sqrt{3}\lambda_c$. In the swirl cases, we observed that, for both the oppose (negative $z_d$) and co-flow (positive $z_d$) configurations, increasing the Swirl number decreases the maximum wavelength/increases the number of fingers. This indicates that the swirling airflow enhances the stretching factor in the droplet deformation process owing to the development of stronger RT instability as compared to the no-swirl condition. Thus, the correlation factor between the maximum unstable wavelength and the critical wavelength will be different in the swirl conditions and one can not simply use $\lambda_m = \sqrt{3}\lambda_c$. In the case of liquid jet/sheet breakup, \cite{vadivukkarasan2017combined} found that the unstable wavelength/growth rate of disturbances is a complex Bessel function in swirl airstream. In order to calculate $\lambda_c$ one has to incorporate the swirl effect on the number of fingers/nodes. We will return to this point again in the next section.

\begin{table}
\begin{center}
\begin{tabular}{cccccc}
$We$  ~~~& No-breakup ~~~& Vibrational breakup ~~~& Retracting bag breakup ~~~& Bag breakup ~~~\\ \hline
11.40  ~~~&   $0 - 1$  ~~~ & $1 - 2$      ~~~         & $4 - 7$    ~~~           & $6 - 11$  ~~~     \\ 
20.86 ~~~&    $0 - 1$ ~~~  & $1 - 3$   ~~~            & $5 - 10$   ~~~             & $10 - 13$   ~~~   \\
\end{tabular}
\end{center}
\caption{The maximum number of fingers observed in different breakup modes for $We=11.40$ and 20.86. For each value of $We$, $\bar z_d$ is varied to obtain the maximum number of fingers associated with different breakup modes. The rest of the parameters are $Sw=0.82$, $\bar x_d = 0.01$, and $\bar y_d = 0.89$.} \label{T3} 
\end{table}

\begin{figure}
\centering
\includegraphics[width=0.6\textwidth]{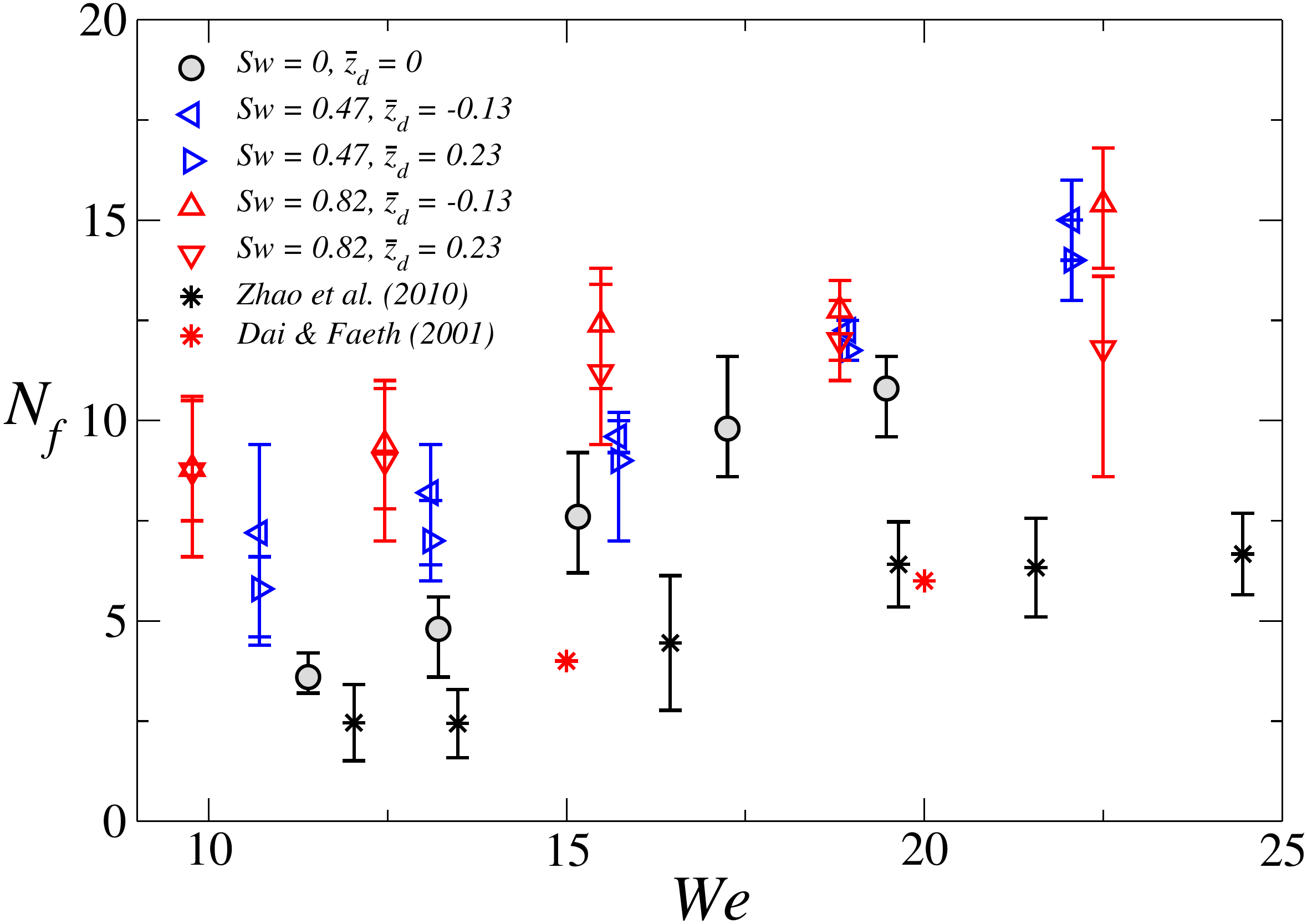}
\caption{Comparison of number of fingers count for different values of Swirl number and position of dispensing needle. The fixed parameters are dispensing needle location in $\bar x_d = 0.01$, and $\bar y_d = 0.89$.}
\label{fig:10}
\end{figure}

Further, we extend this analysis for different breakup modes observed in figure \ref{fig:6} for $We=11.40$ and 20.86. For a fixed $We$, as $\bar z_d$ location of the dispensing needle is shifted from $\bar z_d = -0.63$ to 0.63, the mode changes from no breakup to vibrational breakup to retracting bag breakup to bag breakup mode due to the change in the aerodynamic field and its interaction with the droplet, which in turn increases the number of fingers. Table \ref{T3} presents the total number of fingers observed in different breakup modes for $We=11.40$ and 20.86. The rest of the parameters are $Sw=0.82$, $\bar x_d = 0.01$, and $\bar y_d = 0.89$ (the same as figure \ref{fig:6}). It can be seen that increasing the Weber number (i.e. increasing aerodynamic force) for each mode increases the number of fingers in the majority of cases. The number of fingers also increases as we go from no breakup to bag breakup regions by changing the location of the dispensing needle. The above discussion is summarised in figure \ref{fig:10}, which depicts how the number of fingers $(N_f)$ varies with the Weber number for different values of the Swirl number $(Sw)$ and the $\bar z_d$ location of the dispensing needle. The findings from \cite{zhao2010morphological} and \cite{dai2001temporal} for the no-swirl condition ($Sw=0$, $\bar z_d=0$) are also presented in figure \ref{fig:10}. Comparison of our results to those reported in previous studies for the no-swirl case reveals that while the trend in the variation of $N_f$ is similar, i.e., increasing $We$ increases the value of $N_f$, we observe a higher number of fingers since we have used both views in our analysis as demonstrated in figure \ref{fig:9}. The following observation can be made from figure \ref{fig:10}: (i) increasing the Swirl number for a fixed Weber number increases the number of fingers. (ii) the oppose (negative $\bar z_d$) and co-flow (positive $\bar z_d$) has a negligible effect for each swirl strength.

\subsection{Evolution of drop topology} \label{subsec:evolution}

As discussed in the previous section, the mechanism and growth of the bag created in the swirl flow are different from that in the straight airflow. While the bulging of the disk happens in the direction of the airstream in straight flow, it occurs in the direction of swirling action in swirl flow. In straight airflow, the droplet in the flow field deforms into a disk shape that is perpendicular to the cross-stream direction due to the aerodynamic force. As a result, the disk experiences an unequal pressure distribution across it, bulging outward toward the low-pressure side. The disk then morphs into a bag shape. Subsequently, the interface thins out and eventually bursts because it can no longer withstand the pressure force. The deformation and breakup phenomena in swirl airflow are more complex; the droplet also exhibits a curvilinear trajectory. \cite{kulkarni2014bag,Villermaux2009single} modelled the evolution of the drop topology in the case of straight airflow by considering cross-flow and oppose-flow configurations, respectively. Recently, \cite{jackiw2021aerodynamic} pointed out that the stretching factor, $f$ in cross-flow and oppose-flow configurations will be different, which can have a significant impact on the fragmentation process. Depending on the swirl airstream characteristics, a droplet experiences all cross-flow, oppose-flow, and co-flow configurations in swirl flow. In the following, the earlier approach used in straight airflow has been extended to droplet interaction with a swirling airstream.

A stagnation point flow can be considered for the velocity component of the airstream in the streamwise ($x$) and radial $r (\equiv \sqrt{(y^2+z^2)})$ directions, which are given by
\begin{equation}
U_x = - f \frac{U x}{d_0},   ~  U_r = f \frac{U r}{2 d_0} ,
\end{equation}
where $f$ is the stretching factor, $U_r = \sqrt{U_y^2 + U_z^2}$ denotes the radial velocity and $U$ represents the mean velocity of the airflow field at the exit of the air nozzle. Thus, the resultant velocity of the airstream, $U_{net} = \sqrt{U_x^2+U_r^2}$. This is essentially the same as $U$ as verified by the flow field obtained from our stereo-PIV experiments. 

\begin{figure}
\centering
\includegraphics[width=0.75\textwidth]{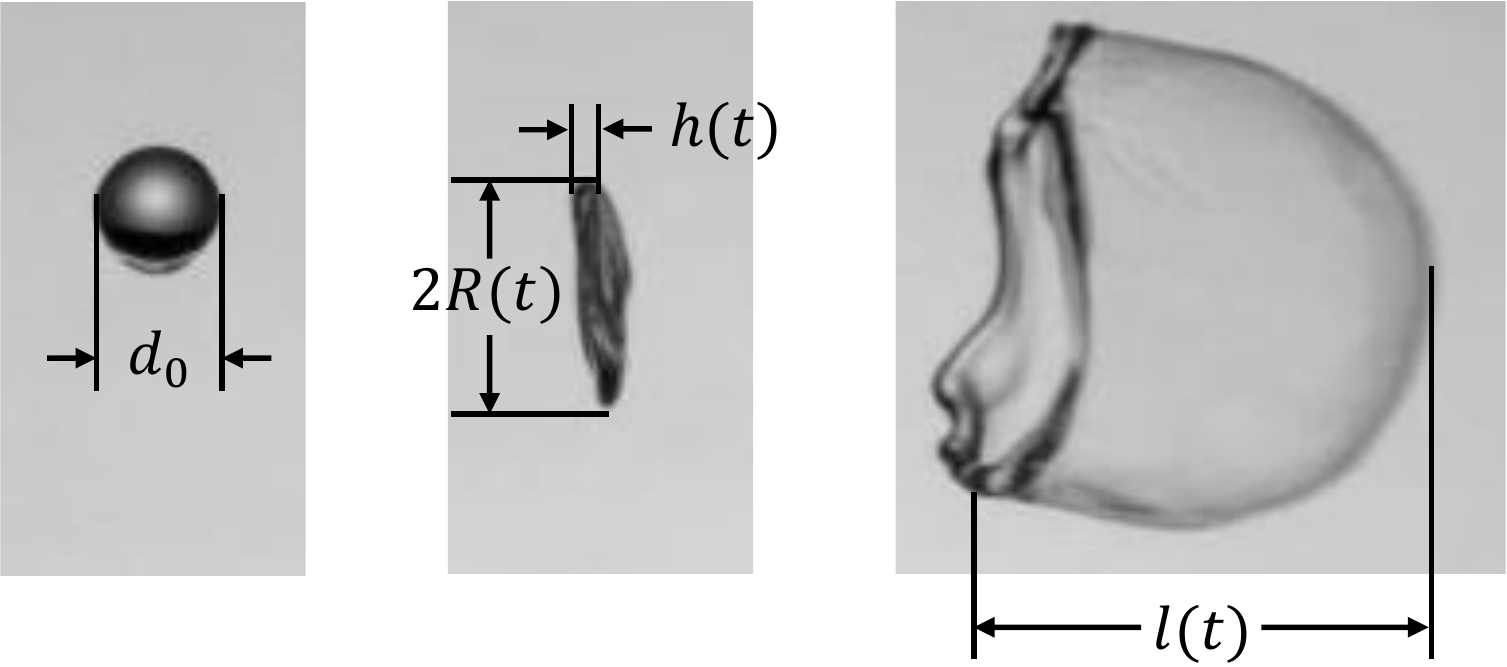}
\caption{Various dimensions of the droplet obtained from the experiments to be used in the modelling. Here, $R(t)$, $l(t)$ and $h(t)$ represent the instantaneous radius, length in the streamwise direction and thickness of the disk of the deformed droplet, respectively.}
\label{fig_dimensions}
\end{figure}

Furthermore, using the Navier-Stokes and continuity equations, the pressure field $(p_a)$ around the droplet can be obtained by assuming the flow is inviscid, incompressible, and quasi-steady \citep{kulkarni2014bag}. 
\begin{equation}
p_a(r,x)=p_a(0) - \rho_a \frac{f^2 U^2}{8 d_0^2} r^2 + \rho_a \frac{f^2 U^2}{2 d_0^2} x^2,
\end{equation}
where $r (\equiv \sqrt{(y^2+z^2)})$ denotes the radial coordinate.  
Thus, at $x=0$:
\begin{equation} \label{eq:pressure}
p_a(r)=p_a(0) - \rho_a \frac{f^2 U^2}{8 d_0^2} r^2.
\end{equation}
Here, $p_a(0) = {\rho_a U^2/2}$ is the stagnation pressure at $(r,x)=(0,0)$.

Assuming a fully developed airstream, the axisymmetric Navier-Stokes equations in the cylindrical coordinate system describe the dynamics of the droplet deforming from a spherical to a disk shape, which is given by
\begin{equation} \label{eq:NS}
\rho_l\left ( \frac{\partial u_r}{\partial t} + u_r \frac{\partial u_r}{\partial r} \right ) = - \frac{\partial p_l}{\partial r} + \mu_l \left [ \frac{1}{r}\frac{\partial}{\partial r} \left ( r \frac{\partial u_r}{\partial r} \right ) - \frac{u_r}{r^2} \right ],
\end{equation}
\begin{equation} \label{mass}
r \frac{\partial h}{\partial t} + u_r \frac{\partial (r u_r h)}{\partial r} = 0 ,
\end{equation}
where $h(t)$ is the time-dependent thickness of the disk (see, figure \ref{fig_dimensions}). As the Reynolds number ($Re \equiv \rho_a U d_0/\mu_a$) in our experiments is in the range of $1420-2130$, assuming the interaction of the droplet with the airstream to be inviscid, the velocity field inside the droplet can be obtained from Eq. (\ref{mass}) as
\begin{equation} \label{velocity}
u_r(r,t) = \frac{r}{R} \left ( \frac{d R}{dt} \right ),
\end{equation}
where $R(t)$ is the radius of the disk shaped droplet (figure \ref{fig_dimensions}). While the tangential stresses are negligible, the normal stress balance at the interface separating the air and liquid phase is given by
\begin{equation} \label{eq:stress}
    \sigma_l \kappa = T_{rr}(l) - T_{rr}(a),  
\end{equation}
where $\kappa$ is the curvature of the interface, $T_{rr}(l) = p_l(r)$ and $T_{rr}(a) = p_a(r)$ are the normal stress components at the interface ($r=R(t)$) associated with the liquid and air phases, respectively. Here, $\kappa = 2/h(t)$. Using Eq. (\ref{eq:pressure}), Eq. (\ref{eq:stress}) becomes
\begin{equation} \label{eq:pressure2}
p_l(R) = p_a(0) - \rho_a {\frac{f^2 U^2}{8 d_0^2} R^2} + \frac{2 \sigma_l}{h}.
\end{equation}
Now we defining, $\alpha(t) = R(t)/(d_0/2)$. Non-dimensionalising and integrating Eq. (\ref{eq:NS}) from $r=0$ to $r=R(t)$ and using Eq. (\ref{eq:pressure2}), we get
\begin{equation} \label{eq:phi2}
    \frac{d^2 \alpha}{d \tau^2} - \left ( \frac{f^2}{4} - \frac{24}{We} \right ) \alpha = 0.
\end{equation}
Solving Eq. (\ref{eq:phi2}) using the initial conditions: $\alpha(0) = 1$ and $\alpha'(0) = 0$, which describe a spherical shape drop during its early interactions with the airstream, we get
\begin{equation} \label{eq:phisol}
\alpha = e^{\tau  \sqrt{\frac{f^2}{4} - \frac{24}{We}}}.
\end{equation}
It can be observed from this equation that, the value of the stretching factor $(f)$ in the cross-flow configuration $(We=12)$ is 2$\sqrt{2}$ \citep{kulkarni2014bag}, whereas $f=4$ in the opposite flow configuration $(We=6)$ \citep{Villermaux2009single}. Thus, we can use our theoretical model to determine the critical Weber number for a range of flow configurations by knowing the stretching factor.

\begin{figure}
\centering
\includegraphics[width=0.6\textwidth]{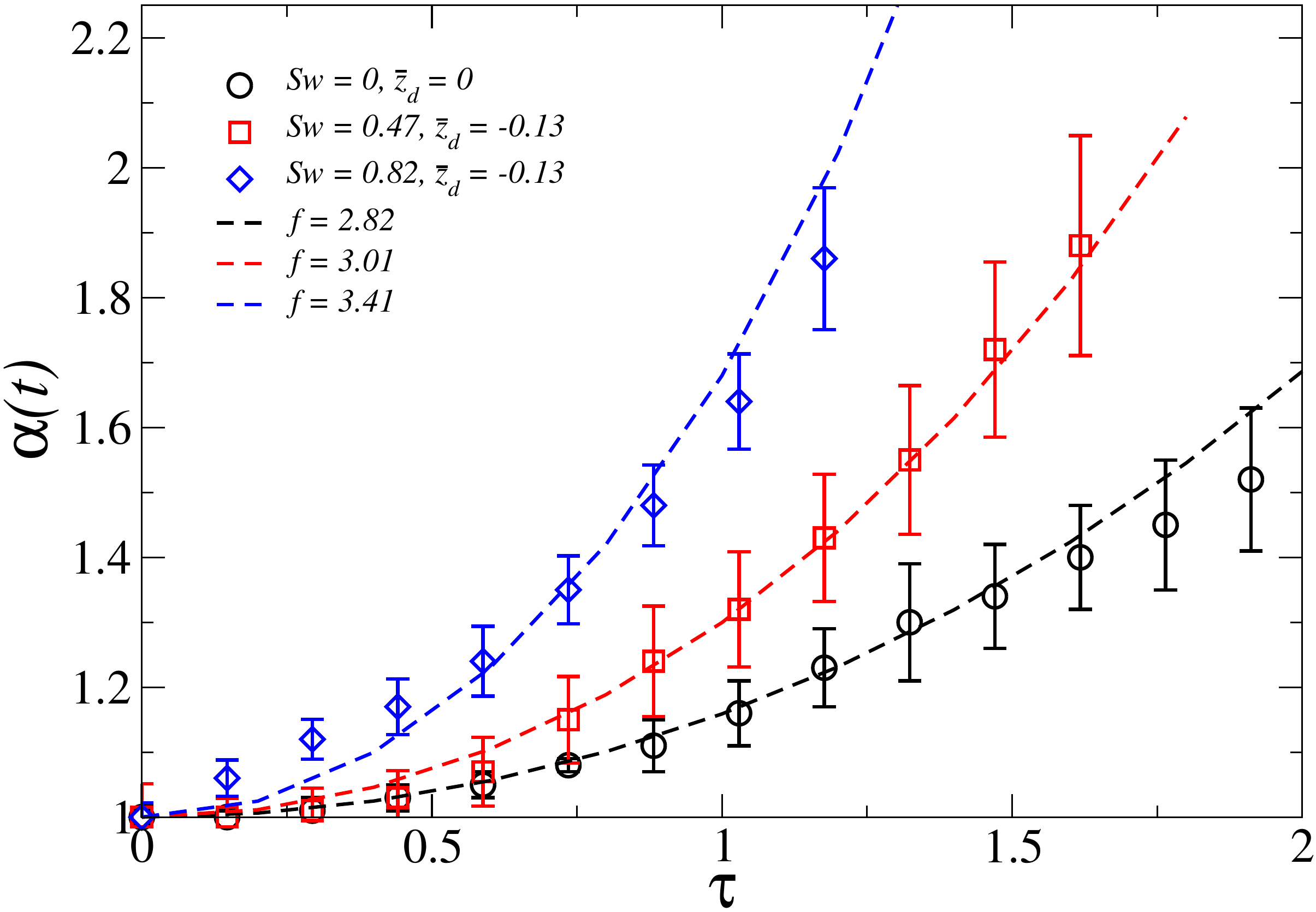}
\caption{Evolution of $\alpha(t)=R(t)/(d_0/2)$ with dimensionless time, $\tau$ for different values of the stretching factor $(f)$ and swirl conditions. Here, $We=15.77$. Symbols represent the experimental results, with an error bar generated from five repetitions. The dashed lines represent the theoretical results.}
\label{fig:11}
\end{figure}

We also found that $\alpha = \cosh \left (\tau  \sqrt{\frac{f^2}{4} - \frac{24}{We}}\right )$ equally accurately predicts the experimental results. Figure \ref{fig:11} depicts the evolution of $\alpha(t)$ with dimensionless time, $\tau$ for different values of the stretching factor $(f)$ and swirl flow conditions. The corresponding theoretical predictions (dashed lines) obtained using the abovementioned equation are also shown for different values of $f$. Our experimental results for the no-swirl case in the cross-flow configuration ($Sw=0, \bar z_d=0$) agrees with the theoretically predicted exponential growth rate of the  droplet evolution using $f=2\sqrt{2} \approx 2.82$ as suggested by \cite{kulkarni2014bag,jackiw2021aerodynamic}. Although we present figure \ref{fig:11} on a linear scale to be consistent with previous studies, the corresponding figure on a logarithmic scale is shown as figure \ref{fig:11_log} (in Appendix) to demonstrate exponential growth of $\alpha(t)$. In swirl cases, the growth rate of $\alpha(t)$ remains exponential, but the slope of the evolution process has increased as the swirl strength has increased, indicating that the swirl flow has expedited the drop deformation process. These findings are also consistent with our analysis of RT instability in the previous section, which demonstrated that the number of fingers increases as the swirl strength increases (see, figure \ref{fig:10}). Also, \cite{soni2020deformation} showed that as the direction of the straight airstream changes from cross-flow to oppose-flow (i.e., by changing the orientation of the interaction of the droplet with airstream), the critical Weber number requirement for bag breakup decreases. It can be seen in figure \ref{fig:11} that the theoretical predictions with $f = 3.01$ and 3.41 agree well with the experimental observations for $Sw=0.47$ and 0.82 at $z_d=-0.13$ (cross-flow configuration), respectively. Thus, it is reasonable to argue that increasing the vane angle (increasing the swirl strength) increases the stretching rate factor and it approaches the value $(f = 4)$ for the perfect opposing flow configuration \citep{Villermaux2009single}. It is also worth noting that the values of the stretching factor, $f$ used in the theoretical predictions in figure \ref{fig:11} are fitted values, but the variation of the stretching factor with increase in the swirl strength is consistent with the velocity fields (figure \ref{fig:3}) obtained from the stereo-PIV measurements. Figure \ref{figR2} (in Appendix) depicts the variation of $U_r/U_x$ at the droplet suspension location ($\bar x_d = 0.01, ~ \bar y_d = 0.89$) with $f$, demonstrating that increasing the swirl strength increases the velocity in the transverse direction, allowing for easier droplet fragmentation.

\begin{figure}
\centering
\includegraphics[width=0.6\textwidth]{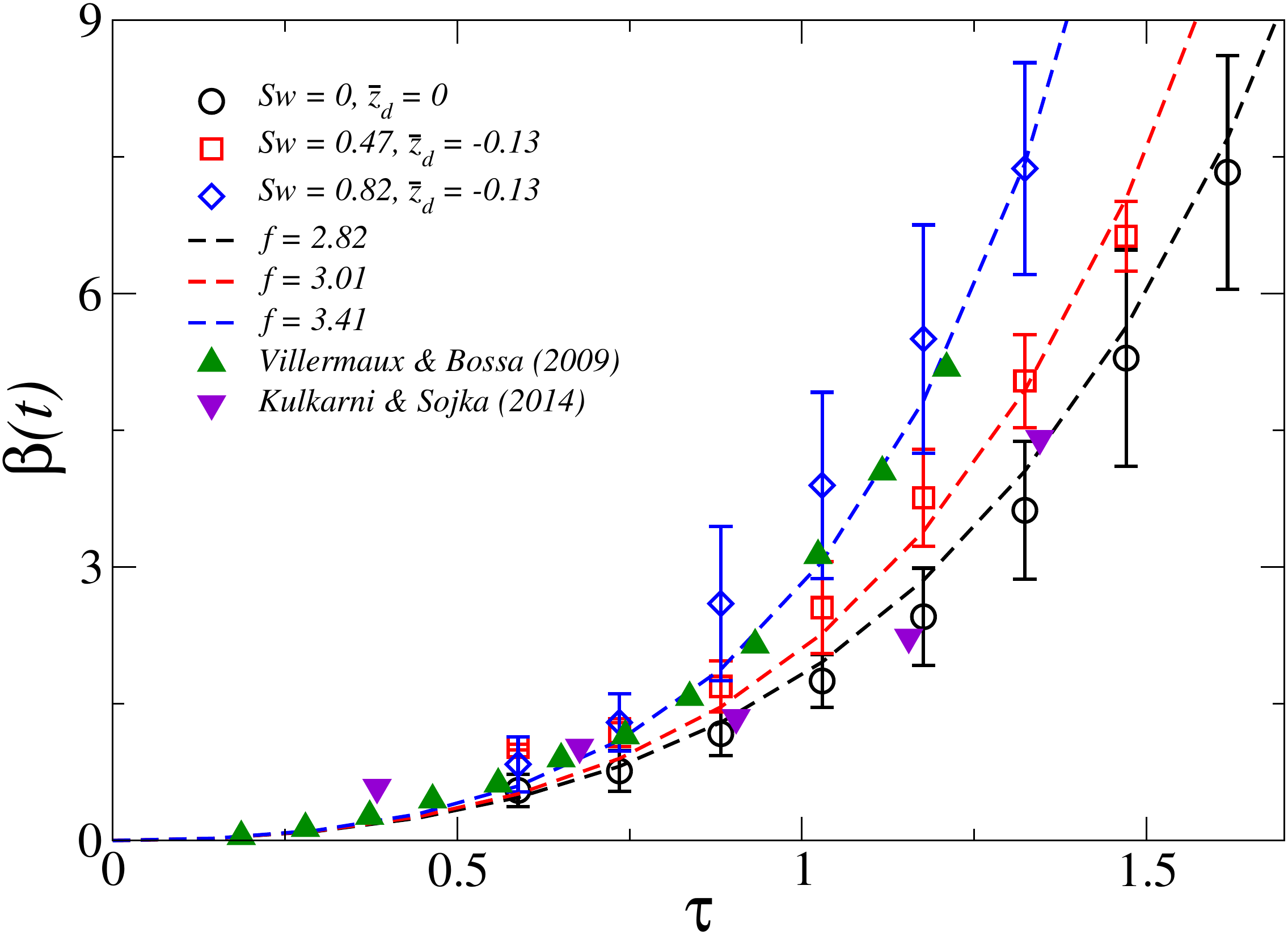}
\caption{Evolution of $\beta(t)=l(t)/(d_0/2)$ with dimensionless time, $\tau$ for different values of the stretching factor $(f)$ and swirl conditions. Here, $We=15.77$. Symbols represent the experimental results, with an error bar generated from five repetitions. The dashed lines represent the theoretical results. The results reported by \cite{kulkarni2014bag} (for cross-flow) and \cite{Villermaux2009single} (oppose-flow) in the case of straight airflow are also shown.}
\label{fig:12}
\end{figure}

Next, we model the growth of the bag and develop its correlation with Weber number, $We$ and stretching rate, $f$. The bag's growth can be described by its elongation, $l(t)$, as shown in figure \ref{fig_dimensions}. For sufficiently large dimensionless time $\tau$, an asymptotic solution of $\alpha= e^{\tau \sqrt{\frac{f^2}{4} - \frac{24}{We}}}$ is used to solve the bag evolution process $\left(\beta(t) = \frac { l(t) } { d_0/2 } \right)$. The differential equation for the elongation of the bag can be obtained by balancing the forces at the tip of the bag \citep{kulkarni2014bag}, which is given by
\begin{equation} \label{eq:beta}
    \frac{d^2 \beta}{d \tau^2} - \frac{24}{We} \beta - 2 \alpha^2 = 0,
\end{equation}
and solved using the initial conditions: $\beta(0) = 0$ and $ \beta'(0) = 0$, which signifies there was no bulging in the disk and no progression of the disk to bulging at the onset of the bag formation. Thus, we get
\begin{equation} \label{eq:betasol}
    \beta = - \frac{1}{\gamma} \left ( \frac{\sqrt{m}}{\sqrt{n}} + 1 \right ) e^{\tau\sqrt{n}} + \frac{1}{\gamma} \left ( \frac{\sqrt{m}}{\sqrt{n}} - 1 \right ) e^{-\tau\sqrt{n}} + \frac{2}{\gamma} e^{\tau \sqrt{m}},
\end{equation}
where $\gamma = f^2 - 120/We$, $m = f^2 - 96/We$ and $n = 24/We$. In figure \ref{fig:12}, we plot the temporal evolution of $\beta$ obtained from our experiments for different swirl conditions. The theoretical predictions obtained using Eq. (\ref{eq:betasol}) are also shown in figure \ref{fig:12} for the corresponding values of the stretching factor, $f$ (as discussed in figure \ref{fig:11}). The experimental results reveal the bag development process displays exponential growth, and match the theoretical predictions quite well. It can also be seen that increasing the swirl strength increases the growth rate of the bag development. We confirm that the theoretical growth rate predictions using $f = 4$ and $3.41$ follow nearly identical trends, as an oblique airstream with obliquity $\ge 45^\circ$ (towards the oppose-flow configuration) produces a critical Weber number $\approx 6$ \citep{soni2020deformation}. This indicates that a swirling airstream interacting with a droplet at an angle greater than $45^\circ$ behaves similar to the oppose-flow configuration. As a consequence, $f = 4$ and $3.41$ perfectly match the results of \cite{Villermaux2009single} (oppose-flow), and $f = 2\sqrt2 $ matches the results of \cite{kulkarni2014bag} (cross-flow) as seen in figure \ref{fig:12}.

\section{Concluding remarks}
\label{sec:conc}
The dynamics of an ethanol droplet interacting with a swirling airflow in an orthogonal arrangement are investigated experimentally and theoretically. The experimental methodology employs a continuous air-jet technique with a mechanism to produce swirl airflow. While a shadowgraphy approach using two high-speed imaging systems is employed to analyze the droplet's trajectory and breakup morphology for different values of swirl strength and Weber number, the particle image velocimetry (PIV) approach is used to examine the flow field created by the imposed swirling airstream from a nozzle. A droplet in a straight airstream mostly experiences no-breakup, vibrational breakup, and bag breakup modes. A new breakup mode, termed as `retracting bag breakup', is found in the swirl flow condition. In this case, as the drop deforms to a disk shape due to its interaction with the airstream, it experiences a differential flow field due to the presence of the wake of vanes and the central recirculation zone in a swirl flow condition in contrast to a drop undergoing convectional breakup modes in straight flow condition. As a result, the bag expands in the upper half of the drop but contracts in the lower half, causing the bag to fragment as it retreats. The influence of the Weber number, Swirl number, and the dispensing needle's location on the droplet dynamics is also studied. A regime map demarcating the various modes is presented for different sets of dimensionless parameters influencing the droplet morphology. The droplet interface becomes unstable due to the Rayleigh-Taylor (RT) instability mechanism, and the perturbations are amplified by the aerodynamic force, resulting in finger-like structures on the rim of the droplet. Compared to the no-swirl scenario, the swirling airflow enhances the stretching factor in the droplet deformation process due to the development of stronger RT instability. The relationship between the maximum wavelength of the undulation on the droplet's surface is strongly dependent on the swirl conditions, and the earlier relationship established for straight airflows cannot be used in the present study. Thus, a modified theoretical analysis based on the Rayleigh-Taylor instability mechanism is proposed for the swirl flow, which agrees well with our experimental results. \\

\noindent{\bf Authors contribution  statement:} P.K.K. and S.K.S. contributed equally to the project; P.K.K. performed the experiments while S.K.S. contributed with the theoretical modelling. All the authors contributed to the analysis of the results and to the preparation of manuscript. The project was formulated and coordinated by K.C.S.\\

\noindent{\bf Declaration of Interests:} The authors report no conflict of interest. \\

\noindent{\bf Acknowledgement:} {K.C.S. thanks the Science \& Engineering Research Board, India for their financial support (Grant number: CRG/2020/000507).}

%\clearpage

\appendix

\section{Supplementary material}

\begin{figure}
\centering
\includegraphics[width=0.7\textwidth]{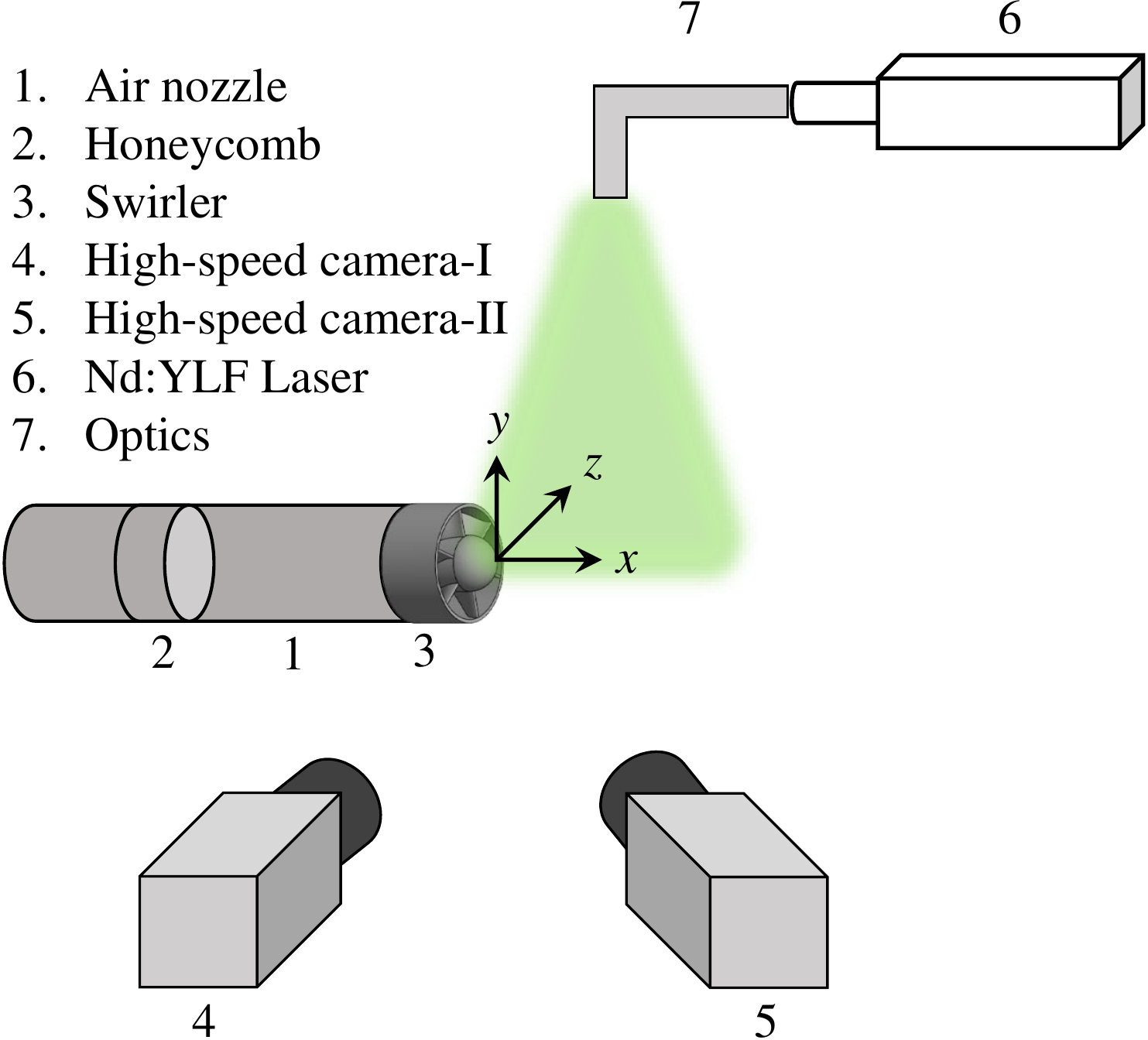}
\caption{Schematic diagram of the stereo-PIV set-up. It consists of two high-speed cameras, an air nozzle with swirler and Nd:YLF laser with sheet optics.}
\label{figS1}
\end{figure}

\begin{figure}
\centering
\includegraphics[height=0.95\textwidth]{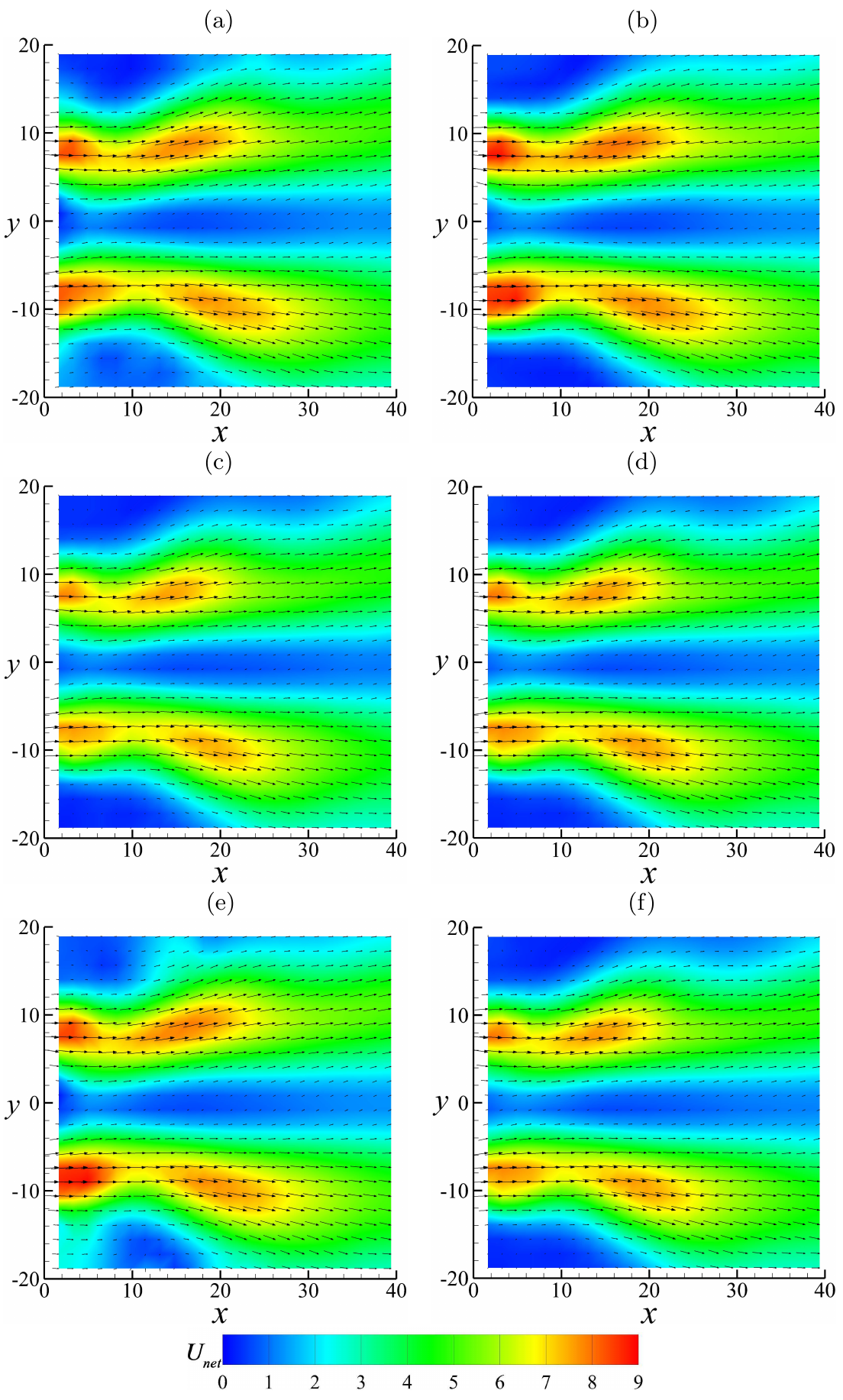} 
\caption{The contours of the resultant velocity, $U_{net} (= \sqrt{U_x^2+U_y^2})$ superimposed with velocity vectors at $\bar z_d = 0$ plane for $We=11.40$ and $Sw=0.82$. The statistical stationary analysis associated with the laser energy intensity of (a) 30\% and (b) 40\% obtained using 1000 frames and 700 Hz; the frame averaging using (c) 500 and (c) 1000 frames for laser energy intensity of 40\% and 700 Hz; the repetition rate (e) 200 Hz and (f) 700 Hz for 40\% and 1000 frames. The $x$ and $y$ labels are in mm and $U_{net}$ is in m/s.}
\label{figS2}
\end{figure}

\clearpage

\begin{figure}
\centering
\includegraphics[width=0.6\textwidth]{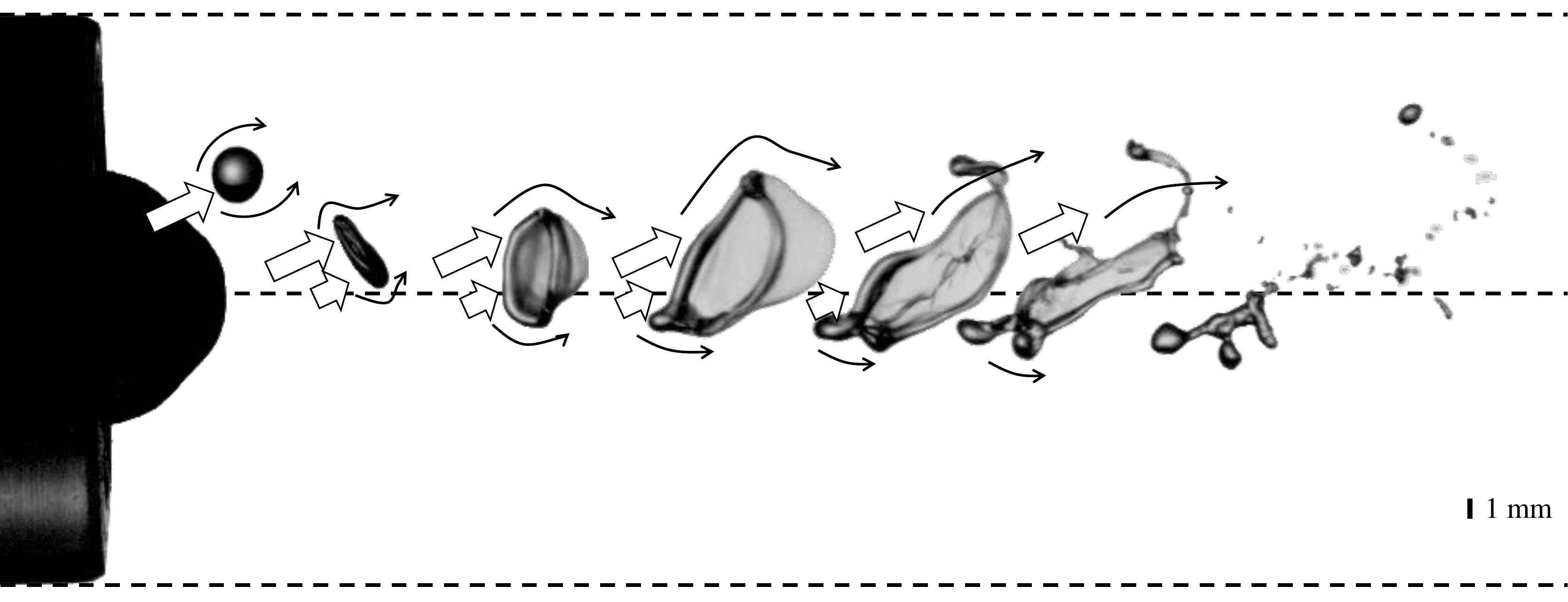}
\caption{The mechanism of the retracting bag breakup morphology of an ethanol droplet (front, $x-y$ view). The parameters are $We = 7.72$, $Sw = 0.82$, $\bar x_d = 0.01$, $\bar y_d = 0.89$ and $\bar z_d = -0.48$. The interaction of the airstream with the droplet is marked by arrow symbols.}
\label{figS4}
\end{figure}

\begin{figure}
\centering
\includegraphics[width=0.6\textwidth]{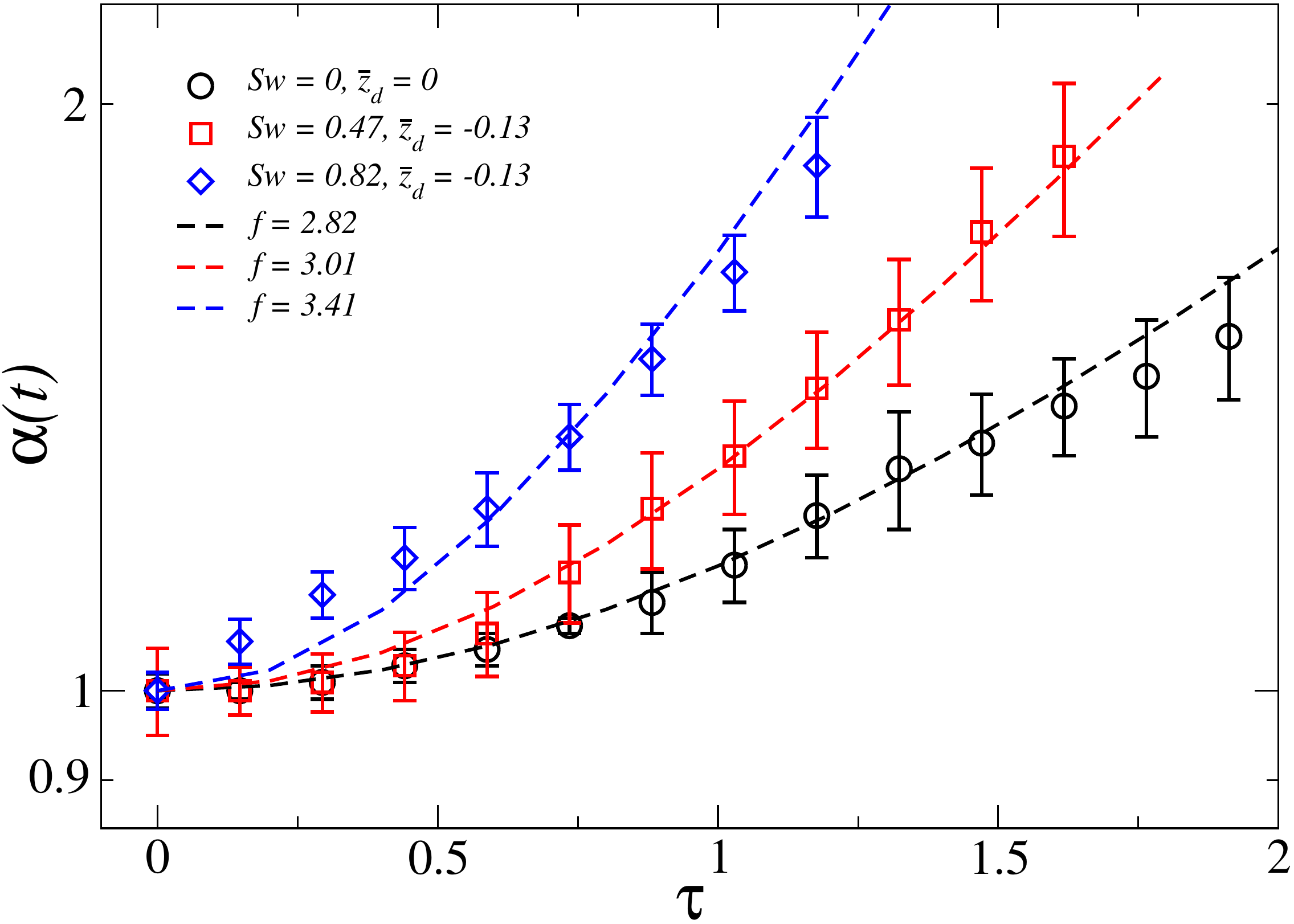}
\caption{The same data as in Figure  \ref{fig:11} on logarithmic scale to demonstrate exponential growth of $\alpha(t)$.}
\label{fig:11_log}
\end{figure}

\begin{figure}
\centering
\includegraphics[width=0.6\textwidth]{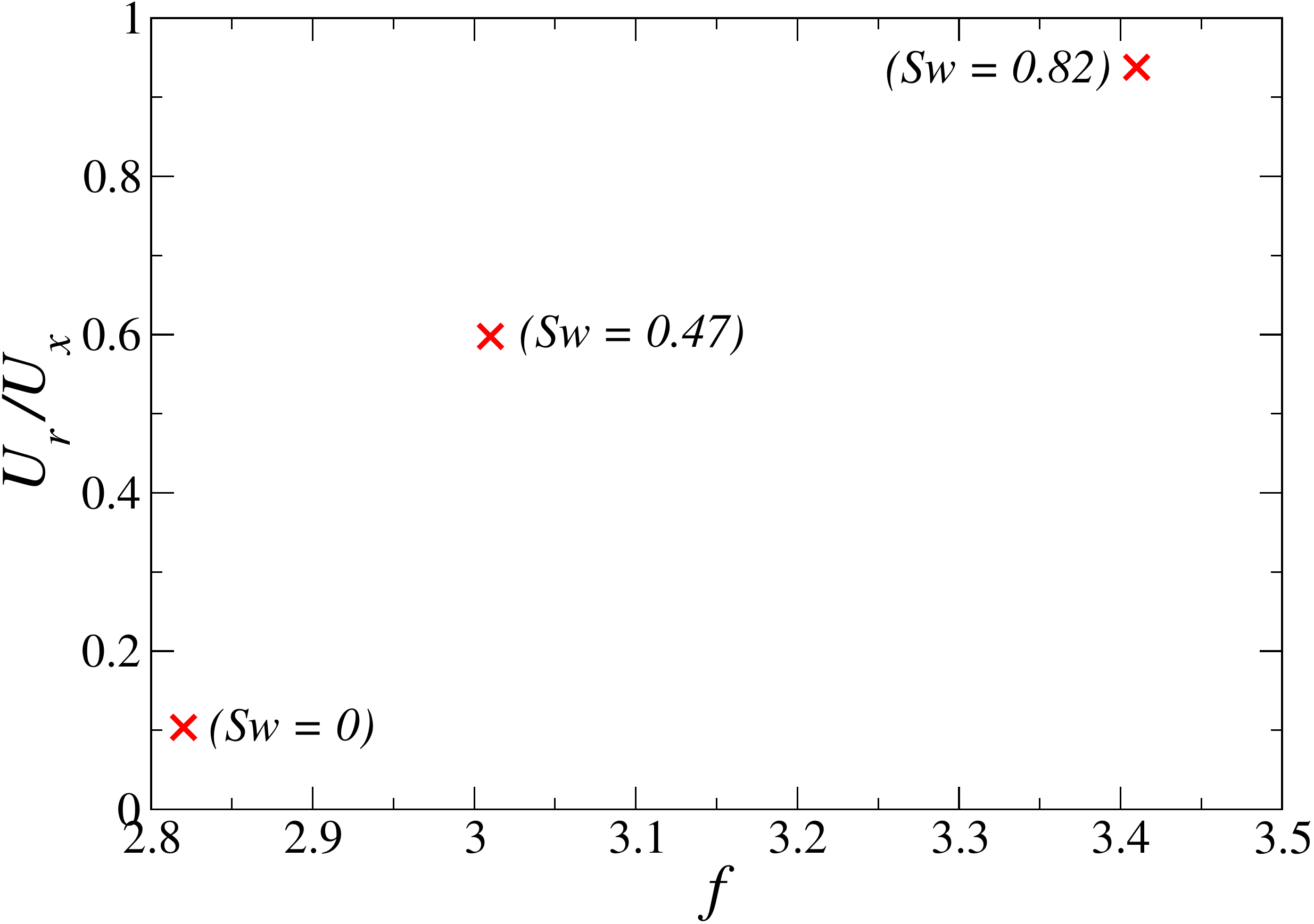}
\caption{Variation of $U_r/U_x$ at the droplet suspension location (${\bar x_d} = 0.01, {\bar y_d} = 0.89$) with the stretching factor, $f$ for different Swirl numbers. Here, $U_r(=\sqrt{U_y^2 + U_z^2})$ and $U_x$ are radial and axial velocity components of the airstream obtained using the Stereo-PIV.}
\label{figR2}
\end{figure}

%\clearpage
%\bibliographystyle{jfm}
%\bibliography{bibl}

\end{document}